\documentclass[aps,prl,floatfix,twocolumn,notitlepage,superscriptaddress,10pt]{revtex4-2}

\usepackage[T1]{fontenc}
\usepackage{xcolor}
\usepackage{grffile}
\usepackage{amsmath,amsthm,amssymb,bbold}
\usepackage{bm}
\usepackage{microtype}
\usepackage{graphicx}
\usepackage{natbib}
\usepackage{dsfont}
\usepackage{hyperref}
\hypersetup{colorlinks,linkcolor=blue,urlcolor=blue,citecolor=red}

\newcommand{\avg}[1]{\left\langle #1\right\rangle}
\newcommand{\one}{\mathds{1}}
\newcommand{\subc}{\mathrm{C}}
\newcommand{\subs}{\mathrm{S}}
\newcommand{\safeincludegraphics}[2][]{%
  \IfFileExists{#2}{\includegraphics[#1]{#2}}{%
    \fbox{\parbox[c][0.17\textheight][c]{0.92\columnwidth}{%
      \centering Insert numerical figure here}}}}

\begin{document}

\title{Dephasing-Induced Universal Spin Fluctuations in a Localized Hubbard Chain}

\author{C\u{a}t\u{a}lin Pa\c{s}cu Moca}
\email{mocap@uoradea.ro}
\affiliation{Department of Physics, University of Oradea, 410087 Oradea, Romania}
\affiliation{Department of Theoretical Physics, Institute of Physics, Budapest University of Technology and Economics, H-1111 Budapest, Hungary}
\affiliation{MTA-BME Lend\"ulet ``Momentum'' Open Quantum Systems Research Group, Budapest University of Technology and Economics, H-1111 Budapest, Hungary}
\author{Ovidiu I. P\^{a}\c{t}u}
\affiliation{Institute for Space Sciences, Bucharest-M\u{a}gurele, R 077125, Romania}
\author{Bal\'azs D\'ora}
\affiliation{Department of Theoretical Physics, Institute of Physics, Budapest University of Technology and Economics, H-1111 Budapest, Hungary}
\affiliation{MTA-BME Lend\"ulet ``Momentum'' Open Quantum Systems Research Group, Budapest University of Technology and Economics, H-1111 Budapest, Hungary}

\begin{abstract}
Decoherence usually washes out quantum effects. We show that in a disordered chain it can instead switch on universal fluctuations. In the infinite-$U$ Hubbard chain, particles cannot pass one another, so the spin crossing a cut is fixed by the number of particles that cross. Without an environment, disorder localizes the charge and the spin fluctuations stop growing.  Dephasing restores diffusion: the spin width grows again as $t^{1/8}$, fastest at an intermediate dephasing rate, and the spin distribution takes the same non-Gaussian M-Wright form as in the clean chain.
\end{abstract}

\maketitle

\paragraph{Introduction.---}
Full counting statistics (FCS) probes an entire transfer distribution and can therefore distinguish 
transport processes that share the same mean or 
variance~\cite{LevitovLesovik1993,LevitovLeeLesovik1996,Klich2003,Schonhammer2007,Schonhammer2009,AbanovIvanov2008,Klich2009,Song2012,DoyonMyers2019,MyersBhaseen2020,Valli2025}. 
Recent works in closed quantum systems have indeed uncovered distinct and sometimes anomalous
distributional classes in diffusive, single-file, and integrable
dynamics~\cite{GopalakrishnanMorningstarVasseurKhemani2024,KrajnikSchmidtPasquierIlievskiProsen2022,KrajnikSchmidtPasquierProsenIlievski2024,
FujimotoIshiyamaKuroseYoshimuraSasamoto2026}.
However, no quantum system is fully isolated, and local monitoring or dephasing can change transport exponents~\cite{GoriniKossakowskiSudarshan1976,Lindblad1976,Prosen2010,Daley2014,Manzano2020,Znidaric2010,Znidaric2011,Prosen2011}. The usual expectation is that decoherence degrades quantum features. In disordered systems the opposite can happen: a random potential localizes noninteracting particles in one dimension~\cite{Anderson1958}, and dephasing destroys the interference responsible for localization, restoring transport~\cite{PlenioHuelga2008,Rebentrost2009}. Here we ask whether decoherence can likewise switch on, rather than destroy, a universal non-Gaussian FCS.

The infinite-$U$ Hubbard chain provides a setting for this question. Double occupation is 
forbidden and particles cannot pass one another. The charge coordinates therefore evolve as spinless 
fermions, while the spatially ordered sequence of spin labels remains frozen~\cite{Hubbard1963,Gutzwiller1963,Kanamori1963,Yang1967,LiebWu1968,OgataShiba1990,EsslerFrahmGohmannKlumperKorepin2005,Kumar2009,ZVR18,ZVR19}. 
The same factorization underlies recent work on strong-coupling spin dynamics~\cite{GamayunHutsalyukPozsgayZvonarev2023,GamayunQuinnBidzhievZvonarev2024} and extends to $SU(N)$ models and cold-atom gases~\cite{Sutherland1975,Schlottmann1987,HonerkampHofstetter2004,Gorshkov2010,CazalillaHoUeda2009,CazalillaRey2014,GuanBatchelorLee2013,CapponiLecheminantTotsuka2016,Taie2012,Pagano2014,Scazza2014,Zhang2014,Hofrichter2016,Kinoshita2006,Rigol2007,Vidmar2016}.

\begin{figure}[!b]
\centering
\safeincludegraphics[width=\columnwidth]{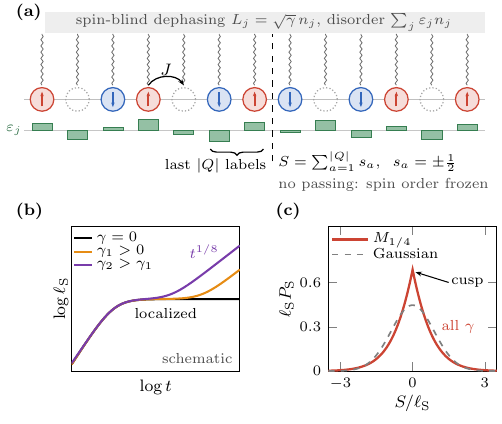}
\caption{Mechanism and main result. (a) Infinite-$U$ chain with 
dephasing and random on-site energies $\varepsilon_j$ (bars). Particles cannot pass one another, so the spin transferred across the 
cut is the sum of the $|Q|$ spin labels nearest to it. By the composition law, Eq.~\eqref{eq:composition}, dephasing and disorder enter only 
through the charge distribution $P_{\subc}$, which fixes the width $\ell_{\subs}$, while the frozen spin order fixes the shape. (b) Schematic 
growth of $\ell_{\subs}=\frac12(\kappa_2^{\subc})^{1/4}$ in a disordered chain. At $\gamma=0$ the charge is Anderson-localized, $\kappa_2^{\subc}
$ saturates, and $\ell_{\subs}$ reaches a plateau set by the localization length. Dephasing restores diffusion, and $\ell_{\subs}$ resumes 
growing as $t^{1/8}$, earlier for stronger dephasing in the weak-dephasing regime. (c) The rescaled spin distribution is the M-Wright function $\mathbb P_{\rm MW}$, Eq.~\eqref{eq:mwright}, for every $\gamma$ and in the diffusive regime of the disordered chain; dashed: Gaussian with the same variance.}
\label{fig:sketch}
\end{figure}
The present question builds on a sequence of clean, constrained-transport results. Krajnik and 
collaborators~\cite{KrajnikSchmidtPasquierIlievskiProsen2022,KrajnikSchmidtPasquierProsenIlievski2024} first derived exact anomalous current fluctuations in a deterministic interacting model 
and subsequently identified universal M-Wright-type fluctuations in charged single-file systems
(see also \cite{KVZ22,GopalakrishnanMorningstarVasseurKhemani2024,GMV24,KSIP24,KIPH25,YK25,YK26,YKBI26,UVGN26,POZS26}).
Stochastic transmission at collisions interpolates between this single-file class and a Gaussian~\cite{KrajnikKlobasBertiniProsen2025}, so the class depends on strict no-passing.
Fujimoto \emph{et al.}~\cite{FujimotoIshiyamaKuroseYoshimuraSasamoto2026} then obtained an exact microscopic M-Wright law for spin transfer in the clean spin-$1/2$ 
infinite-$U$ Hubbard chain. Our recent clean-limit 
work generalized the mechanism to arbitrary $SU(N)$, formulated the exact spin--charge composition 
law, and verified the resulting flavor statistics with matrix-product-state calculations for $N=2,3,$ 
and $4$~\cite{MocaPatuZarandDora2026}. In a closed chain, R\k{e}kas \emph{et al.}~\cite{RekasMierzejewskiLenarcicPrelovsek2026} found that 
integrability-breaking perturbations which keep the spin order frozen make the charge diffusive and the grand-canonical spin transport 
subdiffusive. 

These results concern clean chains. With disorder, the charge, and with it the spin transfer, stops spreading, so no scaling regime exists in 
the absence of an environment. We show that dephasing changes this. The frozen order makes spin transfer a random walk whose number of steps is 
the absolute transferred charge; this exact composition law holds for any disorder and any dephasing strength, so the environment acts only on 
the charge. Without dephasing the spin width saturates at a value set by the localization length. Any $\gamma>0$ restores diffusion, and the 
width grows again as $(Dt)^{1/8}$, with a diffusion constant $D(\gamma,W)$ that is maximal at an intermediate dephasing rate. The rescaled 
distribution is the same M-Wright function~\cite{Wright1933,Mainardi2010} as in the clean chain, where dephasing changes the width growth from $t^{1/4}$ to $t^{1/8}$. The charge variance, which sets the width, is given exactly by the dynamics of a single dephased particle, with or 
without disorder. Fig.~\ref{fig:sketch} summarizes the mechanism and sketches the main result.

\paragraph{Model and exact spin--charge subordination.---}
We consider the projected $SU(2)$ Hubbard chain
\begin{equation}
H_\infty=-J\sum_{j,\sigma}
\left(\widetilde c_{j+1,\sigma}^{\dagger}\widetilde c_{j,\sigma}+\mathrm{H.c.}\right)+\sum_j\varepsilon_j n_j,
\label{eq:H}
\end{equation}
with independent on-site energies $\varepsilon_j$ drawn uniformly from $[-W/2,W/2]$, and with local dephasing jump operators $L_j=\sqrt{\gamma}\,n_j$, where $n_j=n_{j,\uparrow}+n_{j,\downarrow}$.
The density matrix obeys the Gorini--Kossakowski--Sudarshan--Lindblad equation~\cite{GoriniKossakowskiSudarshan1976,Lindblad1976}
\begin{equation}
\dot\rho=-i[H_\infty,\rho]-\frac{\gamma}{2}\sum_j[n_j,[n_j,\rho]].
\label{eq:lindblad}
\end{equation}
We work in the projected infinite-temperature ensemble, in which the local states $|0\rangle$, $|\uparrow\rangle$, and $|\downarrow\rangle$ are equiprobable, the filling is $\bar n=2/3$, and the 
spin labels are independent and unpolarized. Neither the projected hopping, the spin-blind potential, nor the density-dependent 
jump operators can permute these labels, and the ensemble is stationary under Eq.~\eqref{eq:lindblad} for any $W$ and $\gamma$. We set $\hbar=1$; in the figures, $\gamma$ and $W$ are given in units of $J$. Distributions and characteristic functions are written as $P(\cdot,t)$ and $\chi(\lambda,t)$ with their $\gamma$ dependence implicit, whereas cumulants and widths, $\kappa_n(t,\gamma)$ and $\ell_{\subs}(t,\gamma)$, carry it explicitly; their dependence on $W$ is left implicit.

We consider two-time charge and spin transfer across the central cut of a long chain: $Q(t)=N_L(t)-N_L(0)$ and $S(t)=S_L^z(t)-S_L^z(0)$, where $S_L^z=\frac12\sum_{j\in L}(n_{j,\uparrow}-n_{j,\downarrow})$. Their characteristic functions are denoted by $\chi_{\subc}(\lambda,t)$ and $\chi_{\subs}(\lambda,t)$. Conditioned on $Q=m$, the transferred spin is, up to an overall sign, the sum of $|m|$ consecutive independent labels $s_a=\pm1/2$. Hence,  the characteristic function of spin transfer is
\begin{equation}
\chi_{\subs}(\lambda,t\,|\,Q=m)
=\left[\cos\left(\frac{\lambda}{2}\right)\right]^{|m|}.
\end{equation}
Averaging over charge transfer gives the exact composition law 
\begin{equation}
\chi_{\subs}(\lambda,t)
=\sum_{m\in\mathbb Z}P_{\subc}(m,t)
\left[\cos\left(\frac{\lambda}{2}\right)\right]^{|m|}.
\label{eq:composition}
\end{equation}
 with $P_{\subc}(m,t)$ the probability that $m$ particles cross the central cut at time $t$.
Within the specified ensemble, Eq.~\eqref{eq:composition} is exact at all times and for any $\gamma$ and any realization of the disorder. It immediately yields for the second cumulant of spin transfer
\begin{equation}
\kappa_2^{\subs}(t,\gamma)=\frac14\avg{|Q(t)|}.
\label{eq:second_exact}
\end{equation}
Dephasing thus affects spin transfer only through the statistics of $|Q|$. The next moment follows in the same way: each label has fourth cumulant $-1/8$, so
\begin{equation}
\avg{S^4(t)}=\frac{1}{16}\left[3\avg{Q^2(t)}-2\avg{|Q(t)|}\right],
\label{eq:fourth_exact}
\end{equation}
again exact at all times and for any $\gamma$. Equivalently, with $\eta=\cos(\lambda/2)$, one obtains
\begin{equation}
\chi_{\subs}(\lambda,t)=\int_{-\pi}^{\pi}\frac{d\theta}{2\pi}\,
\frac{1-\eta^2}{1-2\eta\cos\theta+\eta^2}\,\chi_{\subc}(\theta,t).
\label{eq:poisson_kernel}
\end{equation}
This exact Poisson-kernel transform of the charge characteristic function cleanly separates the universal spin kernel from the dephasing-dependent charge input.

\paragraph{Charge dynamics.---}
Tracing out the frozen spin labels maps the charge sector exactly onto spinless fermions at filling $\bar n$. With $n_j=f_j^{\dagger}f_j$, the reduced density matrix obeys
\begin{equation}
\begin{aligned}
\dot\rho_{\subc}&=-i[H_{\rm sf},\rho_{\subc}]-\frac{\gamma}{2}\sum_j[n_j,[n_j,\rho_{\subc}]],
\\
H_{\rm sf}&=-J\sum_j(f_{j+1}^{\dagger}f_j+\mathrm{H.c.})+\sum_j\varepsilon_j n_j.
\end{aligned}
\label{eq:charge_reduced}
\end{equation}
This single spinless model connects the unitary, crossover, and Zeno regimes. At $\gamma=0$ and $W=0$, Klich's trace formula gives~\cite{LevitovLeeLesovik1996,Klich2003,Schonhammer2007,Schonhammer2009}
\begin{equation}
\chi_{\subc}^{(0)}(\lambda,t)=
\det\!\left[(1-\bar n)\one+
\bar n e^{-i\lambda P_L}e^{i\lambda P_L(t)}\right],
\label{eq:klich}
\end{equation}
where $P_L$ projects onto the left half of the one-particle Hilbert space. For a half-chain cut in the thermodynamic limit,
\begin{equation}
\kappa_2^{\subc}(t,0)
=\bar n(1-\bar n)\sum_{r=-\infty}^{\infty}|r|J_r^2(2Jt)
\simeq \frac{8}{9\pi}Jt,
\label{eq:variance_unitary}
\end{equation}
where $J_r(x)$ is a Bessel function of the first kind and $\bar n(1-\bar n)=2/9$. The linear growth of the integrated-current second cumulant $\kappa_2^{\subc}$ is the FCS signature of ballistic transport; equivalently, the rms transferred charge grows as $t^{1/2}$~\cite{DoyonMyers2019}.

For $\gamma>0$ the full charge FCS is obtained exactly as a noise average of Klich determinants over a stochastic unraveling of the dephasing, Eq.~\eqref{eq:stochastic_klich} of the Supplemental Material (SM)~\cite{SM}.

\paragraph{Quantum-Zeno limit.---}
For $\gamma\gg J$, adiabatic elimination of bond coherences reduces the charge dynamics to the SSEP with 
hopping rate $D=2J^2/\gamma$, which is also its diffusion constant, as in other dephased one-dimensional chains~\cite{Znidaric2010,Znidaric2011,Prosen2011}. Indeed, the coherence between $|\sigma,0\rangle$ and $|0,\sigma\rangle$ on a bond 
obeys $\dot c=-\gamma c-iJ(w_L-w_R)$, where $w_{L,R}$ are the populations of the two configurations. At equilibrium, the SSEP current-generating function is~\cite{DG09,BodineauDerrida2004,BertiniDeSoleGabrielliJonaLasinioLandim2005,AppertRollandDerridaLecomteVanWijland2008,BertiniDeSoleGabrielliJonaLasinioLandim2015}
\begin{equation}
\ln\chi_{\subc}(\lambda,t)
\simeq \sqrt{D t}\,
F\!\left[-\frac89\sin^2\left(\frac{\lambda}{2}\right)\right],
\label{eq:ssep}
\end{equation}
where
\begin{equation}
F(\omega)=\frac2\pi\int_{0}^{\infty}dk\,
\ln\left(1+\omega e^{-k^2}\right).
\end{equation}
Its small-field expansion gives
\begin{equation}
\kappa_2^{\subc}(t,\gamma\gg J)
\simeq \frac{4}{9\sqrt\pi}
\sqrt{\frac{2J^2t}{\gamma}}=\frac{4}{9\sqrt\pi}\sqrt{Dt}.
\label{eq:variance_zeno}
\end{equation}
For every $\gamma>0$, the numerical data ultimately cross from $\kappa_2^{\subc}\propto t$ to the diffusive law $\kappa_2^{\subc}\propto t^{1/2}$; increasing $\gamma$ brings this crossover to earlier times.
The clean-chain cumulants and their effective exponents are shown in Fig.~\ref{fig:cumulants} of the SM.

\begin{figure}[t]
\centering
\begin{minipage}[t]{0.48\columnwidth}
\raggedright\textbf{(a)}\\
\safeincludegraphics[width=\linewidth]{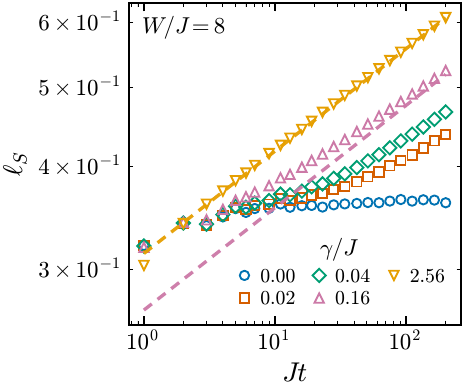}
\end{minipage}\hfill
\begin{minipage}[t]{0.48\columnwidth}
\raggedright\textbf{(b)}\\
\safeincludegraphics[width=\linewidth]{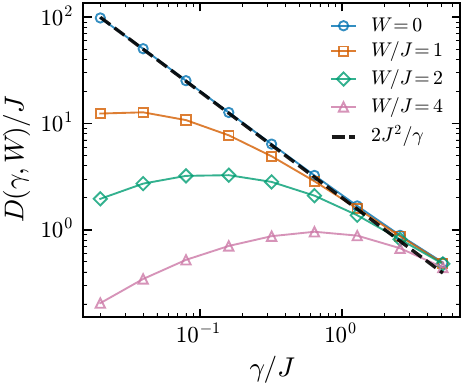}
\end{minipage}
\caption{Disorder and dephasing. (a) Spin width $\ell_{\subs}(t,\gamma)$ for $W=8J$. At $\gamma=0$ the charge is localized and $\ell_{\subs}$ saturates; for $\gamma>0$ it grows again, approaching $t^{1/8}$; dashed: Eq.~\eqref{eq:spin_scale} with $\kappa_2^{\subc}=(4/9\sqrt\pi)\sqrt{Dt}$ and $D$ from (b), for $\gamma=0.16J$ and $2.56J$. (b) Effective diffusion constant $D(\gamma,W)$ versus $\gamma$ for several $W$; dashed: $2J^2/\gamma$. For $W>0$, $D$ is maximal at an intermediate dephasing rate. $L=400$, $Jt\le200$; stochastic Klich determinant averaged over $10^3$ samples, each with its own disorder realization and noise trajectory.}
\label{fig:disorder}
\end{figure}

\paragraph{Exact charge variance at any $\gamma$.---}
The second cumulant does not require the noise average. Because $N_L(t)$ stays quadratic under the adjoint dephasing dynamics, we find (SM) 
\begin{equation}
\kappa_2^{\subc}(t,\gamma)=\bar n(1-\bar n)\sum_{r}|r|\,p(r,t)\equiv\bar n(1-\bar n)\avg{|X(t)|},
\label{eq:kappa2_hsr}
\end{equation}
where $p(r,t)$ is the displacement distribution of a single particle obeying $\dot\varrho=-i[h,\varrho]-\gamma(\varrho-\mathrm{diag}\,\varrho)$, with $h$ the one-particle hopping matrix of $H_{\rm sf}$, i.e., the Haken--Strobl--Reineker model~\cite{HakenReineker1972,HakenStrobl1973}. The derivation does not use translation invariance. With disorder, the same argument gives $\kappa_2^{\subc}=2\bar n(1-\bar n)\sum_{k\in L,\,j\in R}p_t(j|k)$, with $p_t(j|k)$ the transition probability of the dephased particle in the disordered chain, averaged over disorder; a single one-particle evolution started from the left-half projector yields the full curve. At $\gamma=0$, $p(r,t)=J_r^2(2Jt)$ and Eq.~\eqref{eq:kappa2_hsr} reduces to Eq.~\eqref{eq:variance_unitary}. For $\gamma t\gg1$, $X(t)$ is Gaussian with variance $4J^2t/\gamma$, so Eq.~\eqref{eq:kappa2_hsr} reproduces Eq.~\eqref{eq:variance_zeno} for every $\gamma>0$, not only for $\gamma\gg J$; the crossover occurs near $\gamma t\approx\pi/2$.

\paragraph{Disorder: dephasing switches on spin broadening.---}
Figure~\ref{fig:disorder} shows the effect of disorder. At $\gamma=0$ the spinless fermions are Anderson-localized, $\kappa_2^{\subc}$ saturates, and by Eq.~\eqref{eq:spin_scale} the spin width stops growing: for $W=8J$ it levels off at $\ell_{\subs}\approx0.36$ after $Jt\approx5$ [Fig.~\ref{fig:disorder}(a)]. Every $\gamma>0$ restores growth. At long times the charge is diffusive with an effective diffusion constant $D(\gamma,W)$, defined by $\kappa_2^{\subc}\simeq(4/9\sqrt\pi)\sqrt{D t}$, so that $\ell_{\subs}\propto(Dt)^{1/8}$. We obtain $D$ from a linear fit of $[\kappa_2^{\subc}/2\bar n(1-\bar n)]^2=Dt/\pi$ over $100\le Jt\le200$. Figure~\ref{fig:disorder}(b) shows that $D(\gamma,W)$ is non-monotonic for $W>0$. For $\gamma\gtrsim W$ the disorder is irrelevant and $D\simeq2J^2/\gamma$, as in the clean chain. For weak dephasing, $D$ grows with $\gamma$, because each dephasing event lets the particle leave its localized state. The maximum lies at a dephasing rate that increases with $W$. Since $\ell_{\subs}\propto D^{1/8}$ at fixed long $t$, the spin fluctuations are largest at this intermediate decoherence rate.

The typical, central charge fluctuations are asymptotically Gaussian in both the ballistic and diffusive regimes,
\begin{equation}
P_{\subc}(m,t)\simeq
\frac{1}{\sqrt{2\pi \kappa_2^{\subc}(t,\gamma)}}
\exp\!\left[-\frac{m^2}{2\kappa_2^{\subc}(t,\gamma)}\right].
\label{eq:charge_gaussian}
\end{equation}
The full SSEP distribution keeps non-Gaussian corrections from the higher cumulants in Eq.~\eqref{eq:ssep}; the numerical charge distributions collapse onto Eq.~\eqref{eq:charge_gaussian} and follow these corrections. A scalar check of this statement is the ratio $\kappa_4^{\subc}/\kappa_2^{\subc}$, which equals $1-6\bar n(1-\bar n)=-1/3$ for the unitary chain and $1-2\sqrt2/3\approx0.057$ for Eq.~\eqref{eq:ssep}; the numerical ratio reaches the SSEP value for every $\gamma>0$ [Fig.~\ref{fig:charge_distributions}(b) of the SM], so the full long-time charge FCS, not only its variance, is of SSEP type.

\paragraph{M-Wright spin statistics.---}
Substituting the central Gaussian form, Eq.~\eqref{eq:charge_gaussian}, into the exact composition law isolates the scaling of typical spin transfer. With $\phi(\lambda)=-\ln[\cos(\lambda/2)]$, one obtains
\begin{equation}
\chi_{\subs}(\lambda,t)\simeq
\mathcal G\!\left(\sqrt{\kappa_2^{\subc}(t,\gamma)}\,\phi(\lambda)\right),
\label{eq:F_scaling}
\end{equation}
with $\mathcal G(y)=e^{y^2/2}\operatorname{erfc}(y/\sqrt2)$.
Within this central scaling regime, all dephasing dependence enters through $\kappa_2^{\subc}(t,\gamma)$. Since $\phi(\lambda)\simeq\lambda^2/8$, the natural spin counting-field variable is $\lambda\ell_{\subs}$, where
\begin{equation}
\ell_{\subs}(t,\gamma)
=\frac12\left[\kappa_2^{\subc}(t,\gamma)\right]^{1/4}.
\label{eq:spin_scale}
\end{equation}
Thus $\ell_{\subs}\propto t^{1/4}$ for $\gamma=0$, but $\ell_{\subs}\propto t^{1/8}$ for every finite dephasing strength at asymptotically long times.

The probability distribution follows from the same conditioning. At fixed $|Q|=q$, the spin law is exactly binomial, $P_{\subs}(S\,|\,q)=2^{-q}\binom{q}{q/2+S}$.
In the scaling limit, this binomial approaches a Gaussian with variance $q/4$. Mixing it with the half-normal distribution of $|Q|$ gives
\begin{equation}
\!P_{\subs}(S,t)
\simeq \frac{1}{\ell_{\subs}(t,\gamma)}
\mathbb P_{\rm MW}\!\left(\frac{S}{\ell_{\subs}(t,\gamma)}\right),
\label{eq:mwright_scaling}
\end{equation}
where the parameter-free M-Wright scaling function is
\begin{equation}
\mathbb P_{\rm MW}(x)
=\frac1\pi\int_0^\infty\frac{du}{\sqrt u}
\exp\!\left[-\frac{u^2}{2}-\frac{x^2}{2u}\right].
\label{eq:mwright}
\end{equation}
In Mainardi's notation~\cite{Mainardi2010}, $\mathbb P_{\rm MW}(x)=2^{-1/4}M_{1/4}(2^{3/4}|x|)$. It is normalized and has a cusp, $\mathbb P_{\rm MW}(x)\simeq\mathbb P_{\rm MW}(0)-\sqrt{2/\pi}\,|x|$. Its kurtosis is $\avg{x^4}/\avg{x^2}^2=3\pi/2$. 

\begin{figure}[!t]
\centering
\safeincludegraphics[width=0.95\columnwidth]{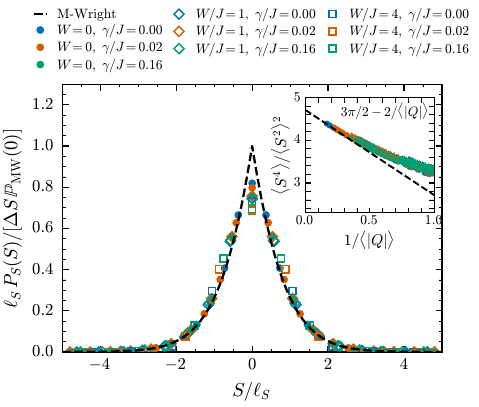}
\caption{Universal central scaling with and without disorder. Rescaled spin distributions $\ell_{\subs}P_{\subs}$, divided by $\Delta S\,\mathbb P_{\rm MW}(0)$ with $\Delta S$ the spacing of spin values, versus $S/\ell_{\subs}$ for $W=0$, $J$, and $4J$ and $\gamma=0$, $0.02J$, and $0.16J$; dashed: M-Wright law, Eq.~\eqref{eq:mwright}. The deficit at the origin reflects the cusp of $\mathbb P_{\rm MW}$ sampled on the lattice of spin values. Inset: spin kurtosis $\avg{S^4}/\avg{S^2}^2$ versus $1/\avg{|Q|}$ for all $W$ and $\gamma$; dashed: $3\pi/2-2/\avg{|Q|}$. $L=400$, $Jt=200$; stochastic Klich determinant averaged over $10^3$ samples, as in Fig.~\ref{fig:disorder}.}
\label{fig:universal_fcs}
\end{figure}

Through Eq.~\eqref{eq:fourth_exact}, the spin kurtosis $\avg{S^4}/\avg{S^2}^2=(3\kappa_2^{\subc}-2\avg{|Q|})/\avg{|Q|}^2$ is known exactly and tends to $3\pi/2$ for every $\gamma$; this single number tests the shape of the distribution independently of its width. For Gaussian charge fluctuations, $\avg{|Q|}^2=2\kappa_2^{\subc}/\pi$, and the exact expression becomes $3\pi/2-2/\avg{|Q|}$: the approach to $3\pi/2$ is slow, controlled by the mean number of crossing particles, and the kurtosis stays below $3\pi/2$ when the charge is localized. 
Plotted against $1/\avg{|Q|}$, all data, clean or disordered, localized or diffusive, should therefore fall on the single parameter-free line $3\pi/2-2/\avg{|Q|}$ [inset of Fig.~\ref{fig:universal_fcs}]. This is a sharper test of universality than the collapse of the distribution itself, which is limited by the small values of $\ell_{\subs}$, and is one of the key predictions of our work.

Equation~\eqref{eq:mwright_scaling} is our central result: dephasing changes the growth law of $\ell_{\subs}$, but not the scaling function. The second cumulant provides a direct test. Combining the exact identity, Eq.~\eqref{eq:second_exact}, with the central Gaussian result $\avg{|Q|}\simeq\sqrt{2\kappa_2^{\subc}/\pi}$ gives
\begin{equation}
\kappa_2^{\subs}(t,\gamma)
\simeq \frac14\sqrt{\frac{2\kappa_2^{\subc}(t,\gamma)}{\pi}}.
\label{eq:spin_variance}
\end{equation}
Thus $\kappa_2^{\subs}\propto t^{1/2}$ at $\gamma=0$, whereas $\kappa_2^{\subs}\propto t^{1/4}$ for $\gamma>0$ at long times, corresponding to the widths $t^{1/4}$ and $t^{1/8}$, respectively.

Equation~\eqref{eq:poisson_kernel} also shows that the spin FCS inherits far-tail information from the full, generally non-Gaussian charge distribution. The M-Wright law corresponds to the central sector of the charge distribution, which is Gaussian in both the ballistic and diffusive regimes. The tails of the spin distribution are therefore sensitive to the large-deviation tails of the charge distribution, which are non-Gaussian in the SSEP regime. The explicit quantum-Zeno spin FCS, obtained by inserting Eq.~\eqref{eq:ssep} into Eq.~\eqref{eq:poisson_kernel}, is given in the SM. 

\paragraph{Numerical tests.---}
We evaluate Eq.~\eqref{eq:stochastic_klich} by evolving the one-particle propagator in a Gaussian white-noise potential and averaging the 
determinants over stochastic trajectories and, for $W>0$, over disorder realizations. This retains the many-body FCS while reducing each realization to a free-fermion calculation. 
For $t\ll\gamma^{-1}$, the charge FCS retains the ballistic structure of the unitary 
chain; at longer times, dephasing suppresses bond coherences and produces diffusive hydrodynamics. Figure~\ref{fig:cumulants}(b) of the SM shows that the 
second spin cumulant follows the exact identity, Eq.~\eqref{eq:second_exact}, throughout this crossover and approaches the asymptote, Eq.~\eqref{eq:spin_variance}. No independent spin fit is required: the measured $\kappa_2^{\subc}$ fixes $\ell_{\subs}$ through Eq.~\eqref{eq:spin_scale}.
The resulting central distributions collapse onto the parameter-free curve, Eq.~\eqref{eq:mwright}, as illustrated in Fig.~\ref{fig:universal_fcs}.
Because $\ell_{\subs}\propto t^{1/8}$ grows slowly, $\ell_{\subs}$ remains of order one lattice unit or smaller in the accessible window for $\gamma>0$, and lattice corrections of relative size $\ell_{\subs}^{-1}$ are visible near the cusp at $x=0$. The same collapse holds for the disordered chains in Fig.~\ref{fig:universal_fcs}. In the localized case $\gamma=0$, the shape still follows from Eq.~\eqref{eq:composition}, but $\ell_{\subs}$ saturates below one lattice spacing, so the continuum M-Wright limit is never reached; dephasing is what lets the distribution broaden into it. Thus the charge sector determines the crossover scale, while the frozen spin sequence 
fixes the scaling function.

\paragraph{Conclusions.---}
In a disordered infinite-$U$ Hubbard chain, decoherence is what makes 
universal spin fluctuations possible. Without dephasing the charge 
localizes and the spin width freezes; dephasing restores diffusion, the 
width grows again as $(Dt)^{1/8}$, fastest at an intermediate decoherence 
rate, and the rescaled distribution is M-Wright. In the clean chain, local 
density dephasing changes the transport exponent while preserving the 
central spin-transfer distribution. 
Frozen particle order makes spin transfer a random sum whose number of 
steps is fixed by charge transfer. Ballistic and diffusive charge dynamics 
therefore yield spin-transfer widths proportional to $t^{1/4}$ and $t^{1/8}$, respectively, but the rescaled central distribution remains M-Wright. 
The far tails retain information about charge large deviations, so the universality concerns typical fluctuations.

The result follows from the exact composition law: dephasing and disorder change the charge statistics but do not scramble the ordered spin labels. This 
separates the crossover scale from the spin scaling function and should also apply to impenetrable multicomponent gases when internal-state order
is preserved. For $SU(N)$ labels~\cite{MocaPatuZarandDora2026} the kernel $[\cos(\lambda/2)]^{|m|}$ is replaced by the $|m|$-th power of the single-label characteristic function~\cite{MocaPatuZarandDora2026}, and Eq.~\eqref{eq:kappa2_hsr} carries over with $\bar n=N/(N+1)$. The stochastic-determinant calculations confirm the crossover and the resulting distributional collapse. Conversely, any process that lets particles exchange order, such as superexchange at finite $U$ or stochastic transmission~\cite{KrajnikKlobasBertiniProsen2025}, breaks Eq.~\eqref{eq:composition}; in a classical automaton, stochastic transmission moves the statistics continuously toward a Gaussian~\cite{KrajnikKlobasBertiniProsen2025}.

Cold-atom microscopy can test these predictions. Spin- and density-resolved Hubbard-chain measurements~\cite{Boll2016,Hilker2017,Vijayan2020}, 
bilayer readout~\cite{Koepsell2020}, and controlled decoherence in an interacting lattice gas~\cite{Bouganne2020} establish the relevant 
capabilities, although they do not directly test the M-Wright law. Disorder can be added with optical speckle or quasiperiodic potentials, and controlled photon scattering provides spin-blind dephasing. Repeated preparations of product states with $|0\rangle$, $|\uparrow\rangle$, and $|\downarrow\rangle$ each occurring with probability $1/3$ on every site, 
followed by spin-resolved snapshots after a spin-independent fluctuating potential, would give both $Q$ and $S$. The experiment can test $4\kappa_2^{\subs}=\avg{|Q|}$, the parameter-free transform in Eq.~\eqref{eq:composition}, and the collapse with $\ell_{\subs}=\tfrac12(\kappa_2^{\subc})^{1/4}$. At finite $U$, this must be done after the dephasing crossover ($t\gtrsim\pi/2\gamma$) but before spin exchange (rate $\sim4J^2/U$, so $t\ll U/4J^2$), doublon production, loss, or boundaries alter the 
frozen-order regime.

\begin{acknowledgments}
This work was supported by the National Research, Development and Innovation Office -- NKFIH Project No.~K142179, 
and by grants of the Ministry of Research, Innovation and Digitization, CNCS/CCCDI-UEFISCDI, under project 
numbers PN-IV-P1-PCE-2023-0159 and PN-IV-P1-PCE-2023-0987. This work was also supported by the HUN-REN Hungarian 
Research Network through the Supported Research Groups Programme, HUN-REN-BME-BCE Quantum Technology Research Group (TKCS-2024/34). O.I.P. acknowledges financial support from Grant No.~30N/2023, provided through the 
National Core Program of the Romanian Ministry of Research, Innovation, and Digitization. We acknowledge the 
Digital Government Development and Project Management Ltd. for access to the Komondor HPC facility, and the 
computing infrastructure of the University of Oradea and IOSIN-PACTES at the Institute of Space Science -- INFLPR Subsidiary, Bucharest-M\u{a}gurele, Romania.

\end{acknowledgments}

\bibliography{references}

\begin{thebibliography}{81}%
\makeatletter
\providecommand \@ifxundefined [1]{%
 \@ifx{#1\undefined}
}%
\providecommand \@ifnum [1]{%
 \ifnum #1\expandafter \@firstoftwo
 \else \expandafter \@secondoftwo
 \fi
}%
\providecommand \@ifx [1]{%
 \ifx #1\expandafter \@firstoftwo
 \else \expandafter \@secondoftwo
 \fi
}%
\providecommand \natexlab [1]{#1}%
\providecommand \enquote  [1]{``#1''}%
\providecommand \bibnamefont  [1]{#1}%
\providecommand \bibfnamefont [1]{#1}%
\providecommand \citenamefont [1]{#1}%
\providecommand \href@noop [0]{\@secondoftwo}%
\providecommand \href [0]{\begingroup \@sanitize@url \@href}%
\providecommand \@href[1]{\@@startlink{#1}\@@href}%
\providecommand \@@href[1]{\endgroup#1\@@endlink}%
\providecommand \@sanitize@url [0]{\catcode `\\12\catcode `\$12\catcode
  `\&12\catcode `\#12\catcode `\^12\catcode `\_12\catcode `\%12\relax}%
\providecommand \@@startlink[1]{}%
\providecommand \@@endlink[0]{}%
\providecommand \url  [0]{\begingroup\@sanitize@url \@url }%
\providecommand \@url [1]{\endgroup\@href {#1}{\urlprefix }}%
\providecommand \urlprefix  [0]{URL }%
\providecommand \Eprint [0]{\href }%
\providecommand \doibase [0]{https://doi.org/}%
\providecommand \selectlanguage [0]{\@gobble}%
\providecommand \bibinfo  [0]{\@secondoftwo}%
\providecommand \bibfield  [0]{\@secondoftwo}%
\providecommand \translation [1]{[#1]}%
\providecommand \BibitemOpen [0]{}%
\providecommand \bibitemStop [0]{}%
\providecommand \bibitemNoStop [0]{.\EOS\space}%
\providecommand \EOS [0]{\spacefactor3000\relax}%
\providecommand \BibitemShut  [1]{\csname bibitem#1\endcsname}%
\let\auto@bib@innerbib\@empty
\bibitem [{\citenamefont {Levitov}\ and\ \citenamefont
  {Lesovik}(1993)}]{LevitovLesovik1993}%
  \BibitemOpen
  \bibfield  {author} {\bibinfo {author} {\bibfnamefont {L.~S.}\ \bibnamefont
  {Levitov}}\ and\ \bibinfo {author} {\bibfnamefont {G.~B.}\ \bibnamefont
  {Lesovik}},\ }\bibfield  {title} {\bibinfo {title} {Charge distribution in
  quantum shot noise},\ }\href@noop {} {\bibfield  {journal} {\bibinfo
  {journal} {JETP Letters}\ }\textbf {\bibinfo {volume} {58}},\ \bibinfo
  {pages} {230} (\bibinfo {year} {1993})}\BibitemShut {NoStop}%
\bibitem [{\citenamefont {Levitov}\ \emph {et~al.}(1996)\citenamefont
  {Levitov}, \citenamefont {Lee},\ and\ \citenamefont
  {Lesovik}}]{LevitovLeeLesovik1996}%
  \BibitemOpen
  \bibfield  {author} {\bibinfo {author} {\bibfnamefont {L.~S.}\ \bibnamefont
  {Levitov}}, \bibinfo {author} {\bibfnamefont {H.}~\bibnamefont {Lee}},\ and\
  \bibinfo {author} {\bibfnamefont {G.~B.}\ \bibnamefont {Lesovik}},\
  }\bibfield  {title} {\bibinfo {title} {Electron counting statistics and
  coherent states of electric current},\ }\href
  {https://doi.org/10.1063/1.531672} {\bibfield  {journal} {\bibinfo  {journal}
  {Journal of Mathematical Physics}\ }\textbf {\bibinfo {volume} {37}},\
  \bibinfo {pages} {4845} (\bibinfo {year} {1996})}\BibitemShut {NoStop}%
\bibitem [{\citenamefont {Klich}(2003)}]{Klich2003}%
  \BibitemOpen
  \bibfield  {author} {\bibinfo {author} {\bibfnamefont {I.}~\bibnamefont
  {Klich}},\ }\bibfield  {title} {\bibinfo {title} {Full counting statistics:
  An elementary derivation of {L}evitov's formula},\ }in\ \href
  {https://doi.org/10.1007/978-94-010-0030-7_17} {\emph {\bibinfo {booktitle}
  {Quantum Noise in Mesoscopic Physics}}},\ \bibinfo {series} {NATO Science
  Series II: Mathematics, Physics and Chemistry}, Vol.~\bibinfo {volume} {97},\
  \bibinfo {editor} {edited by\ \bibinfo {editor} {\bibfnamefont {Y.~V.}\
  \bibnamefont {Nazarov}}}\ (\bibinfo  {publisher} {Springer},\ \bibinfo
  {address} {Dordrecht},\ \bibinfo {year} {2003})\ pp.\ \bibinfo {pages}
  {397--402}\BibitemShut {NoStop}%
\bibitem [{\citenamefont {Sch\"{o}nhammer}(2007)}]{Schonhammer2007}%
  \BibitemOpen
  \bibfield  {author} {\bibinfo {author} {\bibfnamefont {K.}~\bibnamefont
  {Sch\"{o}nhammer}},\ }\bibfield  {title} {\bibinfo {title} {Full counting
  statistics for noninteracting fermions: Exact results and the levitov-lesovik
  formula},\ }\href {https://doi.org/10.1103/PhysRevB.75.205329} {\bibfield
  {journal} {\bibinfo  {journal} {Physical Review B}\ }\textbf {\bibinfo
  {volume} {75}},\ \bibinfo {pages} {205329} (\bibinfo {year}
  {2007})}\BibitemShut {NoStop}%
\bibitem [{\citenamefont {Sch\"{o}nhammer}(2009)}]{Schonhammer2009}%
  \BibitemOpen
  \bibfield  {author} {\bibinfo {author} {\bibfnamefont {K.}~\bibnamefont
  {Sch\"{o}nhammer}},\ }\bibfield  {title} {\bibinfo {title} {Full counting
  statistics for noninteracting fermions: exact finite-temperature results and
  generalized long-time approximation},\ }\href
  {https://doi.org/10.1088/0953-8984/21/49/495306} {\bibfield  {journal}
  {\bibinfo  {journal} {Journal of Physics: Condensed Matter}\ }\textbf
  {\bibinfo {volume} {21}},\ \bibinfo {pages} {495306} (\bibinfo {year}
  {2009})}\BibitemShut {NoStop}%
\bibitem [{\citenamefont {Abanov}\ and\ \citenamefont
  {Ivanov}(2008)}]{AbanovIvanov2008}%
  \BibitemOpen
  \bibfield  {author} {\bibinfo {author} {\bibfnamefont {A.~G.}\ \bibnamefont
  {Abanov}}\ and\ \bibinfo {author} {\bibfnamefont {D.~A.}\ \bibnamefont
  {Ivanov}},\ }\bibfield  {title} {\bibinfo {title} {Allowed charge transfers
  between coherent conductors driven by a time-dependent scatterer},\ }\href
  {https://doi.org/10.1103/PhysRevLett.100.086602} {\bibfield  {journal}
  {\bibinfo  {journal} {Physical Review Letters}\ }\textbf {\bibinfo {volume}
  {100}},\ \bibinfo {pages} {086602} (\bibinfo {year} {2008})}\BibitemShut
  {NoStop}%
\bibitem [{\citenamefont {Klich}\ and\ \citenamefont
  {Levitov}(2009)}]{Klich2009}%
  \BibitemOpen
  \bibfield  {author} {\bibinfo {author} {\bibfnamefont {I.}~\bibnamefont
  {Klich}}\ and\ \bibinfo {author} {\bibfnamefont {L.}~\bibnamefont
  {Levitov}},\ }\bibfield  {title} {\bibinfo {title} {Quantum noise as an
  entanglement meter},\ }\href {https://doi.org/10.1103/PhysRevLett.102.100502}
  {\bibfield  {journal} {\bibinfo  {journal} {Physical Review Letters}\
  }\textbf {\bibinfo {volume} {102}},\ \bibinfo {pages} {100502} (\bibinfo
  {year} {2009})}\BibitemShut {NoStop}%
\bibitem [{\citenamefont {Song}\ \emph {et~al.}(2012)\citenamefont {Song},
  \citenamefont {Rachel}, \citenamefont {Flindt}, \citenamefont {Klich},
  \citenamefont {Laflorencie},\ and\ \citenamefont {Hur}}]{Song2012}%
  \BibitemOpen
  \bibfield  {author} {\bibinfo {author} {\bibfnamefont {H.~F.}\ \bibnamefont
  {Song}}, \bibinfo {author} {\bibfnamefont {S.}~\bibnamefont {Rachel}},
  \bibinfo {author} {\bibfnamefont {C.}~\bibnamefont {Flindt}}, \bibinfo
  {author} {\bibfnamefont {I.}~\bibnamefont {Klich}}, \bibinfo {author}
  {\bibfnamefont {N.}~\bibnamefont {Laflorencie}},\ and\ \bibinfo {author}
  {\bibfnamefont {K.~L.}\ \bibnamefont {Hur}},\ }\bibfield  {title} {\bibinfo
  {title} {Bipartite fluctuations as a probe of many-body entanglement},\
  }\href {https://doi.org/10.1103/PhysRevB.85.035409} {\bibfield  {journal}
  {\bibinfo  {journal} {Physical Review B}\ }\textbf {\bibinfo {volume} {85}},\
  \bibinfo {pages} {035409} (\bibinfo {year} {2012})}\BibitemShut {NoStop}%
\bibitem [{\citenamefont {Doyon}\ and\ \citenamefont
  {Myers}(2020)}]{DoyonMyers2019}%
  \BibitemOpen
  \bibfield  {author} {\bibinfo {author} {\bibfnamefont {B.}~\bibnamefont
  {Doyon}}\ and\ \bibinfo {author} {\bibfnamefont {J.}~\bibnamefont {Myers}},\
  }\bibfield  {title} {\bibinfo {title} {Fluctuations in ballistic transport
  from euler hydrodynamics},\ }\href
  {https://doi.org/10.1007/s00023-019-00860-w} {\bibfield  {journal} {\bibinfo
  {journal} {Annales Henri Poincar{\'e}}\ }\textbf {\bibinfo {volume} {21}},\
  \bibinfo {pages} {255} (\bibinfo {year} {2020})}\BibitemShut {NoStop}%
\bibitem [{\citenamefont {Myers}\ and\ \citenamefont
  {Bhaseen}(2020)}]{MyersBhaseen2020}%
  \BibitemOpen
  \bibfield  {author} {\bibinfo {author} {\bibfnamefont {J.}~\bibnamefont
  {Myers}}\ and\ \bibinfo {author} {\bibfnamefont {M.~J.}\ \bibnamefont
  {Bhaseen}},\ }\bibfield  {title} {\bibinfo {title} {Full counting statistics
  of charge transfer in one-dimensional quantum chains},\ }\href
  {https://doi.org/10.1103/PhysRevB.101.125123} {\bibfield  {journal} {\bibinfo
   {journal} {Physical Review B}\ }\textbf {\bibinfo {volume} {101}},\ \bibinfo
  {pages} {125123} (\bibinfo {year} {2020})}\BibitemShut {NoStop}%
\bibitem [{\citenamefont {Valli}\ \emph {et~al.}(2025)\citenamefont {Valli},
  \citenamefont {Moca}, \citenamefont {Werner}, \citenamefont {Kormos},
  \citenamefont {Krajnik}, \citenamefont {Prosen},\ and\ \citenamefont
  {Zar{\'a}nd}}]{Valli2025}%
  \BibitemOpen
  \bibfield  {author} {\bibinfo {author} {\bibfnamefont {A.}~\bibnamefont
  {Valli}}, \bibinfo {author} {\bibfnamefont {C.~P.}\ \bibnamefont {Moca}},
  \bibinfo {author} {\bibfnamefont {M.~A.}\ \bibnamefont {Werner}}, \bibinfo
  {author} {\bibfnamefont {M.}~\bibnamefont {Kormos}}, \bibinfo {author}
  {\bibfnamefont {{\v Z}.}~\bibnamefont {Krajnik}}, \bibinfo {author}
  {\bibfnamefont {T.}~\bibnamefont {Prosen}},\ and\ \bibinfo {author}
  {\bibfnamefont {G.}~\bibnamefont {Zar{\'a}nd}},\ }\bibfield  {title}
  {\bibinfo {title} {Efficient computation of cumulant evolution and full
  counting statistics: Application to infinite-temperature quantum spin
  chains},\ }\href {https://doi.org/10.1103/PhysRevLett.135.100401} {\bibfield
  {journal} {\bibinfo  {journal} {Physical Review Letters}\ }\textbf {\bibinfo
  {volume} {135}},\ \bibinfo {pages} {100401} (\bibinfo {year}
  {2025})}\BibitemShut {NoStop}%
\bibitem [{\citenamefont {Gopalakrishnan}\ \emph
  {et~al.}(2024{\natexlab{a}})\citenamefont {Gopalakrishnan}, \citenamefont
  {Morningstar}, \citenamefont {Vasseur},\ and\ \citenamefont
  {Khemani}}]{GopalakrishnanMorningstarVasseurKhemani2024}%
  \BibitemOpen
  \bibfield  {author} {\bibinfo {author} {\bibfnamefont {S.}~\bibnamefont
  {Gopalakrishnan}}, \bibinfo {author} {\bibfnamefont {A.}~\bibnamefont
  {Morningstar}}, \bibinfo {author} {\bibfnamefont {R.}~\bibnamefont
  {Vasseur}},\ and\ \bibinfo {author} {\bibfnamefont {V.}~\bibnamefont
  {Khemani}},\ }\bibfield  {title} {\bibinfo {title} {Distinct universality
  classes of diffusive transport from full counting statistics},\ }\href
  {https://doi.org/10.1103/PhysRevB.109.024417} {\bibfield  {journal} {\bibinfo
   {journal} {Physical Review B}\ }\textbf {\bibinfo {volume} {109}},\ \bibinfo
  {pages} {024417} (\bibinfo {year} {2024}{\natexlab{a}})}\BibitemShut
  {NoStop}%
\bibitem [{\citenamefont {Krajnik}\ \emph {et~al.}(2022)\citenamefont
  {Krajnik}, \citenamefont {Schmidt}, \citenamefont {Pasquier}, \citenamefont
  {Ilievski},\ and\ \citenamefont
  {Prosen}}]{KrajnikSchmidtPasquierIlievskiProsen2022}%
  \BibitemOpen
  \bibfield  {author} {\bibinfo {author} {\bibfnamefont {{\v Z}.}~\bibnamefont
  {Krajnik}}, \bibinfo {author} {\bibfnamefont {J.}~\bibnamefont {Schmidt}},
  \bibinfo {author} {\bibfnamefont {V.}~\bibnamefont {Pasquier}}, \bibinfo
  {author} {\bibfnamefont {E.}~\bibnamefont {Ilievski}},\ and\ \bibinfo
  {author} {\bibfnamefont {T.}~\bibnamefont {Prosen}},\ }\bibfield  {title}
  {\bibinfo {title} {Exact anomalous current fluctuations in a deterministic
  interacting model},\ }\href {https://doi.org/10.1103/PhysRevLett.128.160601}
  {\bibfield  {journal} {\bibinfo  {journal} {Physical Review Letters}\
  }\textbf {\bibinfo {volume} {128}},\ \bibinfo {pages} {160601} (\bibinfo
  {year} {2022})}\BibitemShut {NoStop}%
\bibitem [{\citenamefont {Krajnik}\ \emph
  {et~al.}(2024{\natexlab{a}})\citenamefont {Krajnik}, \citenamefont {Schmidt},
  \citenamefont {Pasquier}, \citenamefont {Prosen},\ and\ \citenamefont
  {Ilievski}}]{KrajnikSchmidtPasquierProsenIlievski2024}%
  \BibitemOpen
  \bibfield  {author} {\bibinfo {author} {\bibfnamefont {{\v Z}.}~\bibnamefont
  {Krajnik}}, \bibinfo {author} {\bibfnamefont {J.}~\bibnamefont {Schmidt}},
  \bibinfo {author} {\bibfnamefont {V.}~\bibnamefont {Pasquier}}, \bibinfo
  {author} {\bibfnamefont {T.}~\bibnamefont {Prosen}},\ and\ \bibinfo {author}
  {\bibfnamefont {E.}~\bibnamefont {Ilievski}},\ }\bibfield  {title} {\bibinfo
  {title} {Universal anomalous fluctuations in charged single-file systems},\
  }\href {https://doi.org/10.1103/PhysRevResearch.6.013260} {\bibfield
  {journal} {\bibinfo  {journal} {Physical Review Research}\ }\textbf {\bibinfo
  {volume} {6}},\ \bibinfo {pages} {013260} (\bibinfo {year}
  {2024}{\natexlab{a}})}\BibitemShut {NoStop}%
\bibitem [{\citenamefont {Fujimoto}\ \emph {et~al.}(2026)\citenamefont
  {Fujimoto}, \citenamefont {Ishiyama}, \citenamefont {Kurose}, \citenamefont
  {Yoshimura},\ and\ \citenamefont
  {Sasamoto}}]{FujimotoIshiyamaKuroseYoshimuraSasamoto2026}%
  \BibitemOpen
  \bibfield  {author} {\bibinfo {author} {\bibfnamefont {K.}~\bibnamefont
  {Fujimoto}}, \bibinfo {author} {\bibfnamefont {T.}~\bibnamefont {Ishiyama}},
  \bibinfo {author} {\bibfnamefont {T.}~\bibnamefont {Kurose}}, \bibinfo
  {author} {\bibfnamefont {T.}~\bibnamefont {Yoshimura}},\ and\ \bibinfo
  {author} {\bibfnamefont {T.}~\bibnamefont {Sasamoto}},\ }\href
  {https://doi.org/10.48550/arXiv.2602.24008} {\bibinfo {title} {Exact
  anomalous current fluctuations in quantum many-body dynamics}} (\bibinfo
  {year} {2026}),\ \Eprint {https://arxiv.org/abs/2602.24008} {arXiv:2602.24008
  [cond-mat.stat-mech]} \BibitemShut {NoStop}%
\bibitem [{\citenamefont {Gorini}\ \emph {et~al.}(1976)\citenamefont {Gorini},
  \citenamefont {Kossakowski},\ and\ \citenamefont
  {Sudarshan}}]{GoriniKossakowskiSudarshan1976}%
  \BibitemOpen
  \bibfield  {author} {\bibinfo {author} {\bibfnamefont {V.}~\bibnamefont
  {Gorini}}, \bibinfo {author} {\bibfnamefont {A.}~\bibnamefont
  {Kossakowski}},\ and\ \bibinfo {author} {\bibfnamefont {E.~C.~G.}\
  \bibnamefont {Sudarshan}},\ }\bibfield  {title} {\bibinfo {title} {Completely
  positive dynamical semigroups of {N}-level systems},\ }\href
  {https://doi.org/10.1063/1.522979} {\bibfield  {journal} {\bibinfo  {journal}
  {Journal of Mathematical Physics}\ }\textbf {\bibinfo {volume} {17}},\
  \bibinfo {pages} {821} (\bibinfo {year} {1976})}\BibitemShut {NoStop}%
\bibitem [{\citenamefont {Lindblad}(1976)}]{Lindblad1976}%
  \BibitemOpen
  \bibfield  {author} {\bibinfo {author} {\bibfnamefont {G.}~\bibnamefont
  {Lindblad}},\ }\bibfield  {title} {\bibinfo {title} {On the generators of
  quantum dynamical semigroups},\ }\href {https://doi.org/10.1007/BF01608499}
  {\bibfield  {journal} {\bibinfo  {journal} {Communications in Mathematical
  Physics}\ }\textbf {\bibinfo {volume} {48}},\ \bibinfo {pages} {119}
  (\bibinfo {year} {1976})}\BibitemShut {NoStop}%
\bibitem [{\citenamefont {Prosen}(2010)}]{Prosen2010}%
  \BibitemOpen
  \bibfield  {author} {\bibinfo {author} {\bibfnamefont {T.}~\bibnamefont
  {Prosen}},\ }\bibfield  {title} {\bibinfo {title} {Spectral theorem for the
  lindblad equation for quadratic open fermionic systems},\ }\href
  {https://doi.org/10.1088/1742-5468/2010/07/P07020} {\bibfield  {journal}
  {\bibinfo  {journal} {Journal of Statistical Mechanics: Theory and
  Experiment}\ }\textbf {\bibinfo {volume} {2010}},\ \bibinfo {pages} {P07020}
  (\bibinfo {year} {2010})}\BibitemShut {NoStop}%
\bibitem [{\citenamefont {Daley}(2014)}]{Daley2014}%
  \BibitemOpen
  \bibfield  {author} {\bibinfo {author} {\bibfnamefont {A.~J.}\ \bibnamefont
  {Daley}},\ }\bibfield  {title} {\bibinfo {title} {Quantum trajectories and
  open many-body quantum systems},\ }\href
  {https://doi.org/10.1080/00018732.2014.933502} {\bibfield  {journal}
  {\bibinfo  {journal} {Advances in Physics}\ }\textbf {\bibinfo {volume}
  {63}},\ \bibinfo {pages} {77} (\bibinfo {year} {2014})}\BibitemShut {NoStop}%
\bibitem [{\citenamefont {Manzano}(2020)}]{Manzano2020}%
  \BibitemOpen
  \bibfield  {author} {\bibinfo {author} {\bibfnamefont {D.}~\bibnamefont
  {Manzano}},\ }\bibfield  {title} {\bibinfo {title} {A short introduction to
  the {L}indblad master equation},\ }\href {https://doi.org/10.1063/1.5115323}
  {\bibfield  {journal} {\bibinfo  {journal} {AIP Advances}\ }\textbf {\bibinfo
  {volume} {10}},\ \bibinfo {pages} {025106} (\bibinfo {year}
  {2020})}\BibitemShut {NoStop}%
\bibitem [{\citenamefont {Znidaric}(2010)}]{Znidaric2010}%
  \BibitemOpen
  \bibfield  {author} {\bibinfo {author} {\bibfnamefont {M.}~\bibnamefont
  {Znidaric}},\ }\bibfield  {title} {\bibinfo {title} {Exact solution for a
  diffusive nonequilibrium steady state of an open quantum chain},\ }\href
  {https://doi.org/10.1088/1742-5468/2010/05/L05002} {\bibfield  {journal}
  {\bibinfo  {journal} {Journal of Statistical Mechanics: Theory and
  Experiment}\ }\textbf {\bibinfo {volume} {2010}},\ \bibinfo {pages} {L05002}
  (\bibinfo {year} {2010})}\BibitemShut {NoStop}%
\bibitem [{\citenamefont {Znidaric}(2011)}]{Znidaric2011}%
  \BibitemOpen
  \bibfield  {author} {\bibinfo {author} {\bibfnamefont {M.}~\bibnamefont
  {Znidaric}},\ }\bibfield  {title} {\bibinfo {title} {Spin transport in a
  one-dimensional anisotropic heisenberg model},\ }\href
  {https://doi.org/10.1103/PhysRevLett.106.220601} {\bibfield  {journal}
  {\bibinfo  {journal} {Physical Review Letters}\ }\textbf {\bibinfo {volume}
  {106}},\ \bibinfo {pages} {220601} (\bibinfo {year} {2011})}\BibitemShut
  {NoStop}%
\bibitem [{\citenamefont {Prosen}(2011)}]{Prosen2011}%
  \BibitemOpen
  \bibfield  {author} {\bibinfo {author} {\bibfnamefont {T.}~\bibnamefont
  {Prosen}},\ }\bibfield  {title} {\bibinfo {title} {Exact nonequilibrium
  steady state of a strongly driven open {XXZ} chain},\ }\href
  {https://doi.org/10.1103/PhysRevLett.107.137201} {\bibfield  {journal}
  {\bibinfo  {journal} {Physical Review Letters}\ }\textbf {\bibinfo {volume}
  {107}},\ \bibinfo {pages} {137201} (\bibinfo {year} {2011})}\BibitemShut
  {NoStop}%
\bibitem [{\citenamefont {Anderson}(1958)}]{Anderson1958}%
  \BibitemOpen
  \bibfield  {author} {\bibinfo {author} {\bibfnamefont {P.~W.}\ \bibnamefont
  {Anderson}},\ }\bibfield  {title} {\bibinfo {title} {Absence of diffusion in
  certain random lattices},\ }\href {https://doi.org/10.1103/PhysRev.109.1492}
  {\bibfield  {journal} {\bibinfo  {journal} {Phys. Rev.}\ }\textbf {\bibinfo
  {volume} {109}},\ \bibinfo {pages} {1492} (\bibinfo {year}
  {1958})}\BibitemShut {NoStop}%
\bibitem [{\citenamefont {Plenio}\ and\ \citenamefont
  {Huelga}(2008)}]{PlenioHuelga2008}%
  \BibitemOpen
  \bibfield  {author} {\bibinfo {author} {\bibfnamefont {M.~B.}\ \bibnamefont
  {Plenio}}\ and\ \bibinfo {author} {\bibfnamefont {S.~F.}\ \bibnamefont
  {Huelga}},\ }\bibfield  {title} {\bibinfo {title} {Dephasing-assisted
  transport: quantum networks and biomolecules},\ }\href
  {https://doi.org/10.1088/1367-2630/10/11/113019} {\bibfield  {journal}
  {\bibinfo  {journal} {New J. Phys.}\ }\textbf {\bibinfo {volume} {10}},\
  \bibinfo {pages} {113019} (\bibinfo {year} {2008})}\BibitemShut {NoStop}%
\bibitem [{\citenamefont {Rebentrost}\ \emph {et~al.}(2009)\citenamefont
  {Rebentrost}, \citenamefont {Mohseni}, \citenamefont {Kassal}, \citenamefont
  {Lloyd},\ and\ \citenamefont {Aspuru-Guzik}}]{Rebentrost2009}%
  \BibitemOpen
  \bibfield  {author} {\bibinfo {author} {\bibfnamefont {P.}~\bibnamefont
  {Rebentrost}}, \bibinfo {author} {\bibfnamefont {M.}~\bibnamefont {Mohseni}},
  \bibinfo {author} {\bibfnamefont {I.}~\bibnamefont {Kassal}}, \bibinfo
  {author} {\bibfnamefont {S.}~\bibnamefont {Lloyd}},\ and\ \bibinfo {author}
  {\bibfnamefont {A.}~\bibnamefont {Aspuru-Guzik}},\ }\bibfield  {title}
  {\bibinfo {title} {Environment-assisted quantum transport},\ }\href
  {https://doi.org/10.1088/1367-2630/11/3/033003} {\bibfield  {journal}
  {\bibinfo  {journal} {New J. Phys.}\ }\textbf {\bibinfo {volume} {11}},\
  \bibinfo {pages} {033003} (\bibinfo {year} {2009})}\BibitemShut {NoStop}%
\bibitem [{\citenamefont {Hubbard}(1963)}]{Hubbard1963}%
  \BibitemOpen
  \bibfield  {author} {\bibinfo {author} {\bibfnamefont {J.}~\bibnamefont
  {Hubbard}},\ }\bibfield  {title} {\bibinfo {title} {Electron correlations in
  narrow energy bands},\ }\href {https://doi.org/10.1098/rspa.1963.0204}
  {\bibfield  {journal} {\bibinfo  {journal} {Proceedings of the Royal Society
  of London. Series A. Mathematical and Physical Sciences}\ }\textbf {\bibinfo
  {volume} {276}},\ \bibinfo {pages} {238} (\bibinfo {year}
  {1963})}\BibitemShut {NoStop}%
\bibitem [{\citenamefont {Gutzwiller}(1963)}]{Gutzwiller1963}%
  \BibitemOpen
  \bibfield  {author} {\bibinfo {author} {\bibfnamefont {M.~C.}\ \bibnamefont
  {Gutzwiller}},\ }\bibfield  {title} {\bibinfo {title} {Effect of correlation
  on the ferromagnetism of transition metals},\ }\href
  {https://doi.org/10.1103/PhysRevLett.10.159} {\bibfield  {journal} {\bibinfo
  {journal} {Physical Review Letters}\ }\textbf {\bibinfo {volume} {10}},\
  \bibinfo {pages} {159} (\bibinfo {year} {1963})}\BibitemShut {NoStop}%
\bibitem [{\citenamefont {Kanamori}(1963)}]{Kanamori1963}%
  \BibitemOpen
  \bibfield  {author} {\bibinfo {author} {\bibfnamefont {J.}~\bibnamefont
  {Kanamori}},\ }\bibfield  {title} {\bibinfo {title} {Electron correlation and
  ferromagnetism of transition metals},\ }\href
  {https://doi.org/10.1143/PTP.30.275} {\bibfield  {journal} {\bibinfo
  {journal} {Progress of Theoretical Physics}\ }\textbf {\bibinfo {volume}
  {30}},\ \bibinfo {pages} {275} (\bibinfo {year} {1963})}\BibitemShut
  {NoStop}%
\bibitem [{\citenamefont {Yang}(1967)}]{Yang1967}%
  \BibitemOpen
  \bibfield  {author} {\bibinfo {author} {\bibfnamefont {C.~N.}\ \bibnamefont
  {Yang}},\ }\bibfield  {title} {\bibinfo {title} {Some exact results for the
  many-body problem in one dimension with repulsive delta-function
  interaction},\ }\href {https://doi.org/10.1103/PhysRevLett.19.1312}
  {\bibfield  {journal} {\bibinfo  {journal} {Physical Review Letters}\
  }\textbf {\bibinfo {volume} {19}},\ \bibinfo {pages} {1312} (\bibinfo {year}
  {1967})}\BibitemShut {NoStop}%
\bibitem [{\citenamefont {Lieb}\ and\ \citenamefont {Wu}(1968)}]{LiebWu1968}%
  \BibitemOpen
  \bibfield  {author} {\bibinfo {author} {\bibfnamefont {E.~H.}\ \bibnamefont
  {Lieb}}\ and\ \bibinfo {author} {\bibfnamefont {F.~Y.}\ \bibnamefont {Wu}},\
  }\bibfield  {title} {\bibinfo {title} {Absence of mott transition in an exact
  solution of the short-range, one-band model in one dimension},\ }\href
  {https://doi.org/10.1103/PhysRevLett.20.1445} {\bibfield  {journal} {\bibinfo
   {journal} {Physical Review Letters}\ }\textbf {\bibinfo {volume} {20}},\
  \bibinfo {pages} {1445} (\bibinfo {year} {1968})}\BibitemShut {NoStop}%
\bibitem [{\citenamefont {Ogata}\ and\ \citenamefont
  {Shiba}(1990)}]{OgataShiba1990}%
  \BibitemOpen
  \bibfield  {author} {\bibinfo {author} {\bibfnamefont {M.}~\bibnamefont
  {Ogata}}\ and\ \bibinfo {author} {\bibfnamefont {H.}~\bibnamefont {Shiba}},\
  }\bibfield  {title} {\bibinfo {title} {Bethe-ansatz wave function, momentum
  distribution, and spin correlation in the one-dimensional strongly correlated
  hubbard model},\ }\href {https://doi.org/10.1103/PhysRevB.41.2326} {\bibfield
   {journal} {\bibinfo  {journal} {Physical Review B}\ }\textbf {\bibinfo
  {volume} {41}},\ \bibinfo {pages} {2326} (\bibinfo {year}
  {1990})}\BibitemShut {NoStop}%
\bibitem [{\citenamefont {Essler}\ \emph {et~al.}(2005)\citenamefont {Essler},
  \citenamefont {Frahm}, \citenamefont {G{\"o}hmann}, \citenamefont
  {Kl{\"u}mper},\ and\ \citenamefont
  {Korepin}}]{EsslerFrahmGohmannKlumperKorepin2005}%
  \BibitemOpen
  \bibfield  {author} {\bibinfo {author} {\bibfnamefont {F.~H.~L.}\
  \bibnamefont {Essler}}, \bibinfo {author} {\bibfnamefont {H.}~\bibnamefont
  {Frahm}}, \bibinfo {author} {\bibfnamefont {F.}~\bibnamefont {G{\"o}hmann}},
  \bibinfo {author} {\bibfnamefont {A.}~\bibnamefont {Kl{\"u}mper}},\ and\
  \bibinfo {author} {\bibfnamefont {V.~E.}\ \bibnamefont {Korepin}},\ }\href
  {https://doi.org/10.1017/CBO9780511534843} {\emph {\bibinfo {title} {The
  One-Dimensional Hubbard Model}}}\ (\bibinfo  {publisher} {Cambridge
  University Press},\ \bibinfo {address} {Cambridge},\ \bibinfo {year}
  {2005})\BibitemShut {NoStop}%
\bibitem [{\citenamefont {Kumar}(2009)}]{Kumar2009}%
  \BibitemOpen
  \bibfield  {author} {\bibinfo {author} {\bibfnamefont {B.}~\bibnamefont
  {Kumar}},\ }\bibfield  {title} {\bibinfo {title} {Exact solution of the
  infinite-$u$ hubbard problem and other models in one dimension},\ }\href
  {https://doi.org/10.1103/PhysRevB.79.155121} {\bibfield  {journal} {\bibinfo
  {journal} {Physical Review B}\ }\textbf {\bibinfo {volume} {79}},\ \bibinfo
  {pages} {155121} (\bibinfo {year} {2009})}\BibitemShut {NoStop}%
\bibitem [{\citenamefont {Zhang}\ \emph {et~al.}(2018)\citenamefont {Zhang},
  \citenamefont {Vidmar},\ and\ \citenamefont {Rigol}}]{ZVR18}%
  \BibitemOpen
  \bibfield  {author} {\bibinfo {author} {\bibfnamefont {Y.}~\bibnamefont
  {Zhang}}, \bibinfo {author} {\bibfnamefont {L.}~\bibnamefont {Vidmar}},\ and\
  \bibinfo {author} {\bibfnamefont {M.}~\bibnamefont {Rigol}},\ }\bibfield
  {title} {\bibinfo {title} {Impenetrable $\mathrm{SU}(n)$ fermions in
  one-dimensional lattices},\ }\href
  {https://doi.org/10.1103/PhysRevA.98.042129} {\bibfield  {journal} {\bibinfo
  {journal} {Phys. Rev. A}\ }\textbf {\bibinfo {volume} {98}},\ \bibinfo
  {pages} {042129} (\bibinfo {year} {2018})}\BibitemShut {NoStop}%
\bibitem [{\citenamefont {Zhang}\ \emph {et~al.}(2019)\citenamefont {Zhang},
  \citenamefont {Vidmar},\ and\ \citenamefont {Rigol}}]{ZVR19}%
  \BibitemOpen
  \bibfield  {author} {\bibinfo {author} {\bibfnamefont {Y.}~\bibnamefont
  {Zhang}}, \bibinfo {author} {\bibfnamefont {L.}~\bibnamefont {Vidmar}},\ and\
  \bibinfo {author} {\bibfnamefont {M.}~\bibnamefont {Rigol}},\ }\bibfield
  {title} {\bibinfo {title} {Quantum dynamics of impenetrable $\mathrm{SU}(n)$
  fermions in one-dimensional lattices},\ }\href
  {https://doi.org/10.1103/PhysRevA.99.063605} {\bibfield  {journal} {\bibinfo
  {journal} {Phys. Rev. A}\ }\textbf {\bibinfo {volume} {99}},\ \bibinfo
  {pages} {063605} (\bibinfo {year} {2019})}\BibitemShut {NoStop}%
\bibitem [{\citenamefont {Gamayun}\ \emph {et~al.}(2023)\citenamefont
  {Gamayun}, \citenamefont {Hutsalyuk}, \citenamefont {Pozsgay},\ and\
  \citenamefont {Zvonarev}}]{GamayunHutsalyukPozsgayZvonarev2023}%
  \BibitemOpen
  \bibfield  {author} {\bibinfo {author} {\bibfnamefont {O.}~\bibnamefont
  {Gamayun}}, \bibinfo {author} {\bibfnamefont {A.}~\bibnamefont {Hutsalyuk}},
  \bibinfo {author} {\bibfnamefont {B.}~\bibnamefont {Pozsgay}},\ and\ \bibinfo
  {author} {\bibfnamefont {M.~B.}\ \bibnamefont {Zvonarev}},\ }\bibfield
  {title} {\bibinfo {title} {Finite-temperature spin diffusion in the hubbard
  model in the strong-coupling limit},\ }\href
  {https://doi.org/10.21468/SciPostPhys.15.2.073} {\bibfield  {journal}
  {\bibinfo  {journal} {SciPost Physics}\ }\textbf {\bibinfo {volume} {15}},\
  \bibinfo {pages} {073} (\bibinfo {year} {2023})}\BibitemShut {NoStop}%
\bibitem [{\citenamefont {Gamayun}\ \emph {et~al.}(2024)\citenamefont
  {Gamayun}, \citenamefont {Quinn}, \citenamefont {Bidzhiev},\ and\
  \citenamefont {Zvonarev}}]{GamayunQuinnBidzhievZvonarev2024}%
  \BibitemOpen
  \bibfield  {author} {\bibinfo {author} {\bibfnamefont {O.}~\bibnamefont
  {Gamayun}}, \bibinfo {author} {\bibfnamefont {E.}~\bibnamefont {Quinn}},
  \bibinfo {author} {\bibfnamefont {K.}~\bibnamefont {Bidzhiev}},\ and\
  \bibinfo {author} {\bibfnamefont {M.~B.}\ \bibnamefont {Zvonarev}},\
  }\bibfield  {title} {\bibinfo {title} {Emergence of anyonic correlations from
  spin and charge dynamics in one dimension},\ }\href
  {https://doi.org/10.1103/PhysRevA.109.012209} {\bibfield  {journal} {\bibinfo
   {journal} {Physical Review A}\ }\textbf {\bibinfo {volume} {109}},\ \bibinfo
  {pages} {012209} (\bibinfo {year} {2024})}\BibitemShut {NoStop}%
\bibitem [{\citenamefont {Sutherland}(1975)}]{Sutherland1975}%
  \BibitemOpen
  \bibfield  {author} {\bibinfo {author} {\bibfnamefont {B.}~\bibnamefont
  {Sutherland}},\ }\bibfield  {title} {\bibinfo {title} {Model for a
  multicomponent quantum system},\ }\href
  {https://doi.org/10.1103/PhysRevB.12.3795} {\bibfield  {journal} {\bibinfo
  {journal} {Physical Review B}\ }\textbf {\bibinfo {volume} {12}},\ \bibinfo
  {pages} {3795} (\bibinfo {year} {1975})}\BibitemShut {NoStop}%
\bibitem [{\citenamefont {Schlottmann}(1987)}]{Schlottmann1987}%
  \BibitemOpen
  \bibfield  {author} {\bibinfo {author} {\bibfnamefont {P.}~\bibnamefont
  {Schlottmann}},\ }\bibfield  {title} {\bibinfo {title} {Integrable
  narrow-band model with possible relevance to heavy-fermion systems},\ }\href
  {https://doi.org/10.1103/PhysRevB.36.5177} {\bibfield  {journal} {\bibinfo
  {journal} {Physical Review B}\ }\textbf {\bibinfo {volume} {36}},\ \bibinfo
  {pages} {5177} (\bibinfo {year} {1987})}\BibitemShut {NoStop}%
\bibitem [{\citenamefont {Honerkamp}\ and\ \citenamefont
  {Hofstetter}(2004)}]{HonerkampHofstetter2004}%
  \BibitemOpen
  \bibfield  {author} {\bibinfo {author} {\bibfnamefont {C.}~\bibnamefont
  {Honerkamp}}\ and\ \bibinfo {author} {\bibfnamefont {W.}~\bibnamefont
  {Hofstetter}},\ }\bibfield  {title} {\bibinfo {title} {Ultracold fermions and
  the {SU(N)} hubbard model},\ }\href
  {https://doi.org/10.1103/PhysRevLett.92.170403} {\bibfield  {journal}
  {\bibinfo  {journal} {Physical Review Letters}\ }\textbf {\bibinfo {volume}
  {92}},\ \bibinfo {pages} {170403} (\bibinfo {year} {2004})}\BibitemShut
  {NoStop}%
\bibitem [{\citenamefont {Gorshkov}\ \emph {et~al.}(2010)\citenamefont
  {Gorshkov}, \citenamefont {Hermele}, \citenamefont {Gurarie}, \citenamefont
  {Xu}, \citenamefont {Julienne}, \citenamefont {Ye}, \citenamefont {Zoller},
  \citenamefont {Demler}, \citenamefont {Lukin},\ and\ \citenamefont
  {Rey}}]{Gorshkov2010}%
  \BibitemOpen
  \bibfield  {author} {\bibinfo {author} {\bibfnamefont {A.~V.}\ \bibnamefont
  {Gorshkov}}, \bibinfo {author} {\bibfnamefont {M.}~\bibnamefont {Hermele}},
  \bibinfo {author} {\bibfnamefont {V.}~\bibnamefont {Gurarie}}, \bibinfo
  {author} {\bibfnamefont {C.}~\bibnamefont {Xu}}, \bibinfo {author}
  {\bibfnamefont {P.~S.}\ \bibnamefont {Julienne}}, \bibinfo {author}
  {\bibfnamefont {J.}~\bibnamefont {Ye}}, \bibinfo {author} {\bibfnamefont
  {P.}~\bibnamefont {Zoller}}, \bibinfo {author} {\bibfnamefont
  {E.}~\bibnamefont {Demler}}, \bibinfo {author} {\bibfnamefont {M.~D.}\
  \bibnamefont {Lukin}},\ and\ \bibinfo {author} {\bibfnamefont {A.~M.}\
  \bibnamefont {Rey}},\ }\bibfield  {title} {\bibinfo {title} {Two-orbital
  {SU(N)} magnetism with ultracold alkaline-earth atoms},\ }\href
  {https://doi.org/10.1038/nphys1535} {\bibfield  {journal} {\bibinfo
  {journal} {Nature Physics}\ }\textbf {\bibinfo {volume} {6}},\ \bibinfo
  {pages} {289} (\bibinfo {year} {2010})}\BibitemShut {NoStop}%
\bibitem [{\citenamefont {Cazalilla}\ \emph {et~al.}(2009)\citenamefont
  {Cazalilla}, \citenamefont {Ho},\ and\ \citenamefont
  {Ueda}}]{CazalillaHoUeda2009}%
  \BibitemOpen
  \bibfield  {author} {\bibinfo {author} {\bibfnamefont {M.~A.}\ \bibnamefont
  {Cazalilla}}, \bibinfo {author} {\bibfnamefont {A.~F.}\ \bibnamefont {Ho}},\
  and\ \bibinfo {author} {\bibfnamefont {M.}~\bibnamefont {Ueda}},\ }\bibfield
  {title} {\bibinfo {title} {Ultracold fermi gases with emergent {SU(N)}
  symmetry},\ }\href {https://doi.org/10.1088/1367-2630/11/10/103033}
  {\bibfield  {journal} {\bibinfo  {journal} {New Journal of Physics}\ }\textbf
  {\bibinfo {volume} {11}},\ \bibinfo {pages} {103033} (\bibinfo {year}
  {2009})}\BibitemShut {NoStop}%
\bibitem [{\citenamefont {Cazalilla}\ and\ \citenamefont
  {Rey}(2014)}]{CazalillaRey2014}%
  \BibitemOpen
  \bibfield  {author} {\bibinfo {author} {\bibfnamefont {M.~A.}\ \bibnamefont
  {Cazalilla}}\ and\ \bibinfo {author} {\bibfnamefont {A.~M.}\ \bibnamefont
  {Rey}},\ }\bibfield  {title} {\bibinfo {title} {Ultracold fermi gases with
  emergent {SU(N)} symmetry},\ }\href
  {https://doi.org/10.1088/0034-4885/77/12/124401} {\bibfield  {journal}
  {\bibinfo  {journal} {Reports on Progress in Physics}\ }\textbf {\bibinfo
  {volume} {77}},\ \bibinfo {pages} {124401} (\bibinfo {year}
  {2014})}\BibitemShut {NoStop}%
\bibitem [{\citenamefont {Guan}\ \emph {et~al.}(2013)\citenamefont {Guan},
  \citenamefont {Batchelor},\ and\ \citenamefont {Lee}}]{GuanBatchelorLee2013}%
  \BibitemOpen
  \bibfield  {author} {\bibinfo {author} {\bibfnamefont {X.-W.}\ \bibnamefont
  {Guan}}, \bibinfo {author} {\bibfnamefont {M.~T.}\ \bibnamefont
  {Batchelor}},\ and\ \bibinfo {author} {\bibfnamefont {C.}~\bibnamefont
  {Lee}},\ }\bibfield  {title} {\bibinfo {title} {Fermi gases in one dimension:
  From bethe ansatz to experiments},\ }\href
  {https://doi.org/10.1103/RevModPhys.85.1633} {\bibfield  {journal} {\bibinfo
  {journal} {Reviews of Modern Physics}\ }\textbf {\bibinfo {volume} {85}},\
  \bibinfo {pages} {1633} (\bibinfo {year} {2013})}\BibitemShut {NoStop}%
\bibitem [{\citenamefont {Capponi}\ \emph {et~al.}(2016)\citenamefont
  {Capponi}, \citenamefont {Lecheminant},\ and\ \citenamefont
  {Totsuka}}]{CapponiLecheminantTotsuka2016}%
  \BibitemOpen
  \bibfield  {author} {\bibinfo {author} {\bibfnamefont {S.}~\bibnamefont
  {Capponi}}, \bibinfo {author} {\bibfnamefont {P.}~\bibnamefont
  {Lecheminant}},\ and\ \bibinfo {author} {\bibfnamefont {K.}~\bibnamefont
  {Totsuka}},\ }\bibfield  {title} {\bibinfo {title} {Phases of one-dimensional
  {SU(N)} cold atomic fermi gases: From molecular luttinger liquids to
  topological phases},\ }\href {https://doi.org/10.1016/j.aop.2016.01.013}
  {\bibfield  {journal} {\bibinfo  {journal} {Annals of Physics}\ }\textbf
  {\bibinfo {volume} {367}},\ \bibinfo {pages} {50} (\bibinfo {year}
  {2016})}\BibitemShut {NoStop}%
\bibitem [{\citenamefont {Taie}\ \emph {et~al.}(2012)\citenamefont {Taie},
  \citenamefont {Yamazaki}, \citenamefont {Sugawa},\ and\ \citenamefont
  {Takahashi}}]{Taie2012}%
  \BibitemOpen
  \bibfield  {author} {\bibinfo {author} {\bibfnamefont {S.}~\bibnamefont
  {Taie}}, \bibinfo {author} {\bibfnamefont {R.}~\bibnamefont {Yamazaki}},
  \bibinfo {author} {\bibfnamefont {S.}~\bibnamefont {Sugawa}},\ and\ \bibinfo
  {author} {\bibfnamefont {Y.}~\bibnamefont {Takahashi}},\ }\bibfield  {title}
  {\bibinfo {title} {An {SU}(6) mott insulator of an atomic fermi gas realized
  by large-spin pomeranchuk cooling},\ }\href
  {https://doi.org/10.1038/nphys2430} {\bibfield  {journal} {\bibinfo
  {journal} {Nature Physics}\ }\textbf {\bibinfo {volume} {8}},\ \bibinfo
  {pages} {825} (\bibinfo {year} {2012})}\BibitemShut {NoStop}%
\bibitem [{\citenamefont {Pagano}\ \emph {et~al.}(2014)\citenamefont {Pagano},
  \citenamefont {Mancini}, \citenamefont {Cappellini}, \citenamefont
  {Lombardi}, \citenamefont {Sch{\"a}fer}, \citenamefont {Hu}, \citenamefont
  {Liu}, \citenamefont {Catani}, \citenamefont {Sias}, \citenamefont
  {Inguscio},\ and\ \citenamefont {Fallani}}]{Pagano2014}%
  \BibitemOpen
  \bibfield  {author} {\bibinfo {author} {\bibfnamefont {G.}~\bibnamefont
  {Pagano}}, \bibinfo {author} {\bibfnamefont {M.}~\bibnamefont {Mancini}},
  \bibinfo {author} {\bibfnamefont {G.}~\bibnamefont {Cappellini}}, \bibinfo
  {author} {\bibfnamefont {P.}~\bibnamefont {Lombardi}}, \bibinfo {author}
  {\bibfnamefont {F.}~\bibnamefont {Sch{\"a}fer}}, \bibinfo {author}
  {\bibfnamefont {H.}~\bibnamefont {Hu}}, \bibinfo {author} {\bibfnamefont
  {X.-J.}\ \bibnamefont {Liu}}, \bibinfo {author} {\bibfnamefont
  {J.}~\bibnamefont {Catani}}, \bibinfo {author} {\bibfnamefont
  {C.}~\bibnamefont {Sias}}, \bibinfo {author} {\bibfnamefont {M.}~\bibnamefont
  {Inguscio}},\ and\ \bibinfo {author} {\bibfnamefont {L.}~\bibnamefont
  {Fallani}},\ }\bibfield  {title} {\bibinfo {title} {A one-dimensional liquid
  of fermions with tunable spin},\ }\href {https://doi.org/10.1038/nphys2878}
  {\bibfield  {journal} {\bibinfo  {journal} {Nature Physics}\ }\textbf
  {\bibinfo {volume} {10}},\ \bibinfo {pages} {198} (\bibinfo {year}
  {2014})}\BibitemShut {NoStop}%
\bibitem [{\citenamefont {Scazza}\ \emph {et~al.}(2014)\citenamefont {Scazza},
  \citenamefont {Hofrichter}, \citenamefont {H\"{o}fer}, \citenamefont {Groot},
  \citenamefont {Bloch},\ and\ \citenamefont {F\"{o}lling}}]{Scazza2014}%
  \BibitemOpen
  \bibfield  {author} {\bibinfo {author} {\bibfnamefont {F.}~\bibnamefont
  {Scazza}}, \bibinfo {author} {\bibfnamefont {C.}~\bibnamefont {Hofrichter}},
  \bibinfo {author} {\bibfnamefont {M.}~\bibnamefont {H\"{o}fer}}, \bibinfo
  {author} {\bibfnamefont {P.~C.~D.}\ \bibnamefont {Groot}}, \bibinfo {author}
  {\bibfnamefont {I.}~\bibnamefont {Bloch}},\ and\ \bibinfo {author}
  {\bibfnamefont {S.}~\bibnamefont {F\"{o}lling}},\ }\bibfield  {title}
  {\bibinfo {title} {Observation of two-orbital spin-exchange interactions with
  ultracold {SU(N)}-symmetric fermions},\ }\href
  {https://doi.org/10.1038/nphys3061} {\bibfield  {journal} {\bibinfo
  {journal} {Nature Physics}\ }\textbf {\bibinfo {volume} {10}},\ \bibinfo
  {pages} {779} (\bibinfo {year} {2014})}\BibitemShut {NoStop}%
\bibitem [{\citenamefont {Zhang}\ \emph {et~al.}(2014)\citenamefont {Zhang},
  \citenamefont {Bishof}, \citenamefont {Bromley}, \citenamefont {Kraus},
  \citenamefont {Safronova}, \citenamefont {Zoller}, \citenamefont {Rey},\ and\
  \citenamefont {Ye}}]{Zhang2014}%
  \BibitemOpen
  \bibfield  {author} {\bibinfo {author} {\bibfnamefont {X.}~\bibnamefont
  {Zhang}}, \bibinfo {author} {\bibfnamefont {M.}~\bibnamefont {Bishof}},
  \bibinfo {author} {\bibfnamefont {S.~L.}\ \bibnamefont {Bromley}}, \bibinfo
  {author} {\bibfnamefont {C.~V.}\ \bibnamefont {Kraus}}, \bibinfo {author}
  {\bibfnamefont {M.~S.}\ \bibnamefont {Safronova}}, \bibinfo {author}
  {\bibfnamefont {P.}~\bibnamefont {Zoller}}, \bibinfo {author} {\bibfnamefont
  {A.~M.}\ \bibnamefont {Rey}},\ and\ \bibinfo {author} {\bibfnamefont
  {J.}~\bibnamefont {Ye}},\ }\bibfield  {title} {\bibinfo {title}
  {Spectroscopic observation of {SU(N)}-symmetric interactions in {Sr} orbital
  magnetism},\ }\href {https://doi.org/10.1126/science.1254978} {\bibfield
  {journal} {\bibinfo  {journal} {Science}\ }\textbf {\bibinfo {volume}
  {345}},\ \bibinfo {pages} {1467} (\bibinfo {year} {2014})}\BibitemShut
  {NoStop}%
\bibitem [{\citenamefont {Hofrichter}\ \emph {et~al.}(2016)\citenamefont
  {Hofrichter}, \citenamefont {Riegger}, \citenamefont {Scazza}, \citenamefont
  {H\"{o}fer}, \citenamefont {Fernandes}, \citenamefont {Bloch},\ and\
  \citenamefont {F\"{o}lling}}]{Hofrichter2016}%
  \BibitemOpen
  \bibfield  {author} {\bibinfo {author} {\bibfnamefont {C.}~\bibnamefont
  {Hofrichter}}, \bibinfo {author} {\bibfnamefont {L.}~\bibnamefont {Riegger}},
  \bibinfo {author} {\bibfnamefont {F.}~\bibnamefont {Scazza}}, \bibinfo
  {author} {\bibfnamefont {M.}~\bibnamefont {H\"{o}fer}}, \bibinfo {author}
  {\bibfnamefont {D.~R.}\ \bibnamefont {Fernandes}}, \bibinfo {author}
  {\bibfnamefont {I.}~\bibnamefont {Bloch}},\ and\ \bibinfo {author}
  {\bibfnamefont {S.}~\bibnamefont {F\"{o}lling}},\ }\bibfield  {title}
  {\bibinfo {title} {Direct probing of the mott crossover in the {SU(N)}
  fermi-hubbard model},\ }\href {https://doi.org/10.1103/PhysRevX.6.021030}
  {\bibfield  {journal} {\bibinfo  {journal} {Physical Review X}\ }\textbf
  {\bibinfo {volume} {6}},\ \bibinfo {pages} {021030} (\bibinfo {year}
  {2016})}\BibitemShut {NoStop}%
\bibitem [{\citenamefont {Kinoshita}\ \emph {et~al.}(2006)\citenamefont
  {Kinoshita}, \citenamefont {Wenger},\ and\ \citenamefont
  {Weiss}}]{Kinoshita2006}%
  \BibitemOpen
  \bibfield  {author} {\bibinfo {author} {\bibfnamefont {T.}~\bibnamefont
  {Kinoshita}}, \bibinfo {author} {\bibfnamefont {T.}~\bibnamefont {Wenger}},\
  and\ \bibinfo {author} {\bibfnamefont {D.~S.}\ \bibnamefont {Weiss}},\
  }\bibfield  {title} {\bibinfo {title} {A quantum newton's cradle},\ }\href
  {https://doi.org/10.1038/nature04693} {\bibfield  {journal} {\bibinfo
  {journal} {Nature}\ }\textbf {\bibinfo {volume} {440}},\ \bibinfo {pages}
  {900} (\bibinfo {year} {2006})}\BibitemShut {NoStop}%
\bibitem [{\citenamefont {Rigol}\ \emph {et~al.}(2007)\citenamefont {Rigol},
  \citenamefont {Dunjko}, \citenamefont {Yurovsky},\ and\ \citenamefont
  {Olshanii}}]{Rigol2007}%
  \BibitemOpen
  \bibfield  {author} {\bibinfo {author} {\bibfnamefont {M.}~\bibnamefont
  {Rigol}}, \bibinfo {author} {\bibfnamefont {V.}~\bibnamefont {Dunjko}},
  \bibinfo {author} {\bibfnamefont {V.}~\bibnamefont {Yurovsky}},\ and\
  \bibinfo {author} {\bibfnamefont {M.}~\bibnamefont {Olshanii}},\ }\bibfield
  {title} {\bibinfo {title} {Relaxation in a completely integrable many-body
  quantum system: An {Ab Initio} study of the dynamics of highly excited states
  of 1d lattice hard-core bosons},\ }\href
  {https://doi.org/10.1103/PhysRevLett.98.050405} {\bibfield  {journal}
  {\bibinfo  {journal} {Physical Review Letters}\ }\textbf {\bibinfo {volume}
  {98}},\ \bibinfo {pages} {050405} (\bibinfo {year} {2007})}\BibitemShut
  {NoStop}%
\bibitem [{\citenamefont {Vidmar}\ and\ \citenamefont
  {Rigol}(2016)}]{Vidmar2016}%
  \BibitemOpen
  \bibfield  {author} {\bibinfo {author} {\bibfnamefont {L.}~\bibnamefont
  {Vidmar}}\ and\ \bibinfo {author} {\bibfnamefont {M.}~\bibnamefont {Rigol}},\
  }\bibfield  {title} {\bibinfo {title} {Generalized gibbs ensemble in
  integrable lattice models},\ }\href
  {https://doi.org/10.1088/1742-5468/2016/06/064007} {\bibfield  {journal}
  {\bibinfo  {journal} {Journal of Statistical Mechanics: Theory and
  Experiment}\ }\textbf {\bibinfo {volume} {2016}},\ \bibinfo {pages} {064007}
  (\bibinfo {year} {2016})}\BibitemShut {NoStop}%
\bibitem [{\citenamefont {Kormos}\ \emph {et~al.}(2022)\citenamefont {Kormos},
  \citenamefont {V\"{o}r\"{o}s},\ and\ \citenamefont {Zar\'{a}nd}}]{KVZ22}%
  \BibitemOpen
  \bibfield  {author} {\bibinfo {author} {\bibfnamefont {M.}~\bibnamefont
  {Kormos}}, \bibinfo {author} {\bibfnamefont {D.}~\bibnamefont
  {V\"{o}r\"{o}s}},\ and\ \bibinfo {author} {\bibfnamefont {G.}~\bibnamefont
  {Zar\'{a}nd}},\ }\bibfield  {title} {\bibinfo {title} {{Finite-temperature
  dynamics in gapped one-dimensional models in the sine-Gordon family}},\
  }\href {https://doi.org/10.1103/PhysRevB.106.205151} {\bibfield  {journal}
  {\bibinfo  {journal} {Phys. Rev. B}\ }\textbf {\bibinfo {volume} {106}},\
  \bibinfo {pages} {205151} (\bibinfo {year} {2022})}\BibitemShut {NoStop}%
\bibitem [{\citenamefont {Gopalakrishnan}\ \emph
  {et~al.}(2024{\natexlab{b}})\citenamefont {Gopalakrishnan}, \citenamefont
  {McCulloch},\ and\ \citenamefont {Vasseur}}]{GMV24}%
  \BibitemOpen
  \bibfield  {author} {\bibinfo {author} {\bibfnamefont {S.}~\bibnamefont
  {Gopalakrishnan}}, \bibinfo {author} {\bibfnamefont {E.}~\bibnamefont
  {McCulloch}},\ and\ \bibinfo {author} {\bibfnamefont {R.}~\bibnamefont
  {Vasseur}},\ }\bibfield  {title} {\bibinfo {title} {{Non-Gaussian diffusive
  fluctuations in Dirac fluids}},\ }\href
  {https://doi.org/10.1073/pnas.2403327121} {\bibfield  {journal} {\bibinfo
  {journal} {Proceedings of the National Academy of Sciences}\ }\textbf
  {\bibinfo {volume} {121}},\ \bibinfo {pages} {e2403327121} (\bibinfo {year}
  {2024}{\natexlab{b}})}\BibitemShut {NoStop}%
\bibitem [{\citenamefont {Krajnik}\ \emph
  {et~al.}(2024{\natexlab{b}})\citenamefont {Krajnik}, \citenamefont {Schmidt},
  \citenamefont {Ilievski},\ and\ \citenamefont {Prosen}}]{KSIP24}%
  \BibitemOpen
  \bibfield  {author} {\bibinfo {author} {\bibfnamefont {{\v{Z}}.}~\bibnamefont
  {Krajnik}}, \bibinfo {author} {\bibfnamefont {J.}~\bibnamefont {Schmidt}},
  \bibinfo {author} {\bibfnamefont {E.}~\bibnamefont {Ilievski}},\ and\
  \bibinfo {author} {\bibfnamefont {T.}~\bibnamefont {Prosen}},\ }\bibfield
  {title} {\bibinfo {title} {{Dynamical Criticality of Magnetization Transfer
  in Integrable Spin Chains}},\ }\href
  {https://doi.org/10.1103/PhysRevLett.132.017101} {\bibfield  {journal}
  {\bibinfo  {journal} {Phys. Rev. Lett.}\ }\textbf {\bibinfo {volume} {132}},\
  \bibinfo {pages} {017101} (\bibinfo {year} {2024}{\natexlab{b}})}\BibitemShut
  {NoStop}%
\bibitem [{\citenamefont {Krajnik}\ \emph
  {et~al.}(2025{\natexlab{a}})\citenamefont {Krajnik}, \citenamefont
  {Ilievski}, \citenamefont {Prosen}, \citenamefont {H\'{e}ry},\ and\
  \citenamefont {Pasquier}}]{KIPH25}%
  \BibitemOpen
  \bibfield  {author} {\bibinfo {author} {\bibfnamefont {{\v{Z}}.}~\bibnamefont
  {Krajnik}}, \bibinfo {author} {\bibfnamefont {E.}~\bibnamefont {Ilievski}},
  \bibinfo {author} {\bibfnamefont {T.}~\bibnamefont {Prosen}}, \bibinfo
  {author} {\bibfnamefont {B.~J.~A.}\ \bibnamefont {H\'{e}ry}},\ and\ \bibinfo
  {author} {\bibfnamefont {V.}~\bibnamefont {Pasquier}},\ }\bibfield  {title}
  {\bibinfo {title} {{Integrable fishnet circuits and Brownian solitons}},\
  }\href {https://doi.org/10.21468/SciPostPhys.19.1.027} {\bibfield  {journal}
  {\bibinfo  {journal} {SciPost Phys.}\ }\textbf {\bibinfo {volume} {19}},\
  \bibinfo {pages} {027} (\bibinfo {year} {2025}{\natexlab{a}})}\BibitemShut
  {NoStop}%
\bibitem [{\citenamefont {Yoshimura}\ and\ \citenamefont
  {Krajnik}(2025)}]{YK25}%
  \BibitemOpen
  \bibfield  {author} {\bibinfo {author} {\bibfnamefont {T.}~\bibnamefont
  {Yoshimura}}\ and\ \bibinfo {author} {\bibfnamefont {{\v{Z}}.}~\bibnamefont
  {Krajnik}},\ }\bibfield  {title} {\bibinfo {title} {{Anomalous current
  fluctuations from Euler hydrodynamics}},\ }\href
  {https://doi.org/10.1103/PhysRevE.111.024141} {\bibfield  {journal} {\bibinfo
   {journal} {Phys. Rev. E}\ }\textbf {\bibinfo {volume} {111}},\ \bibinfo
  {pages} {024141} (\bibinfo {year} {2025})}\BibitemShut {NoStop}%
\bibitem [{\citenamefont {Yoshimura}\ and\ \citenamefont
  {Krajnik}(2026)}]{YK26}%
  \BibitemOpen
  \bibfield  {author} {\bibinfo {author} {\bibfnamefont {T.}~\bibnamefont
  {Yoshimura}}\ and\ \bibinfo {author} {\bibfnamefont {{\v{Z}}.}~\bibnamefont
  {Krajnik}},\ }\bibfield  {title} {\bibinfo {title} {{Hydrodynamic
  fluctuations of stochastic charged cellular automata}},\ }\href
  {https://doi.org/10.1103/sjxq-hclt} {\bibfield  {journal} {\bibinfo
  {journal} {Phys. Rev. E}\ }\textbf {\bibinfo {volume} {113}},\ \bibinfo
  {pages} {064120} (\bibinfo {year} {2026})}\BibitemShut {NoStop}%
\bibitem [{\citenamefont {Yoshimura}\ \emph {et~al.}(2026)\citenamefont
  {Yoshimura}, \citenamefont {Krajnik}, \citenamefont {Bastianello},\ and\
  \citenamefont {Ilievski}}]{YKBI26}%
  \BibitemOpen
  \bibfield  {author} {\bibinfo {author} {\bibfnamefont {T.}~\bibnamefont
  {Yoshimura}}, \bibinfo {author} {\bibfnamefont {{\v{Z}}.}~\bibnamefont
  {Krajnik}}, \bibinfo {author} {\bibfnamefont {A.}~\bibnamefont
  {Bastianello}},\ and\ \bibinfo {author} {\bibfnamefont {E.}~\bibnamefont
  {Ilievski}},\ }\href {https://doi.org/10.48550/arXiv.2602.24242} {\bibinfo
  {title} {{Anomalous hydrodynamic fluctuations in the quantum XXZ spin
  chain}}} (\bibinfo {year} {2026})\BibitemShut {NoStop}%
\bibitem [{\citenamefont {Urilyon}\ \emph {et~al.}(2026)\citenamefont
  {Urilyon}, \citenamefont {Vasseur}, \citenamefont {Gopalakrishnan},\ and\
  \citenamefont {De~Nardis}}]{UVGN26}%
  \BibitemOpen
  \bibfield  {author} {\bibinfo {author} {\bibfnamefont {A.}~\bibnamefont
  {Urilyon}}, \bibinfo {author} {\bibfnamefont {R.}~\bibnamefont {Vasseur}},
  \bibinfo {author} {\bibfnamefont {S.}~\bibnamefont {Gopalakrishnan}},\ and\
  \bibinfo {author} {\bibfnamefont {J.}~\bibnamefont {De~Nardis}},\ }\href
  {https://doi.org/10.48550/arXiv.2603.02171} {\bibinfo {title} {{Anomalous
  Diffusion and Superdiffusion in Integrable Spin Chains via a Hard-Rod Gas
  Mapping}}} (\bibinfo {year} {2026})\BibitemShut {NoStop}%
\bibitem [{\citenamefont {Pozsgay}(2026)}]{POZS26}%
  \BibitemOpen
  \bibfield  {author} {\bibinfo {author} {\bibfnamefont {B.}~\bibnamefont
  {Pozsgay}},\ }\href {https://doi.org/10.48550/arXiv.2608.10536} {\bibinfo
  {title} {{Anomalous current fluctuations in the stochastic XNOR hopping
  model}}} (\bibinfo {year} {2026})\BibitemShut {NoStop}%
\bibitem [{\citenamefont {Krajnik}\ \emph
  {et~al.}(2025{\natexlab{b}})\citenamefont {Krajnik}, \citenamefont {Klobas},
  \citenamefont {Bertini},\ and\ \citenamefont
  {Prosen}}]{KrajnikKlobasBertiniProsen2025}%
  \BibitemOpen
  \bibfield  {author} {\bibinfo {author} {\bibfnamefont {{\v Z}.}~\bibnamefont
  {Krajnik}}, \bibinfo {author} {\bibfnamefont {K.}~\bibnamefont {Klobas}},
  \bibinfo {author} {\bibfnamefont {B.}~\bibnamefont {Bertini}},\ and\ \bibinfo
  {author} {\bibfnamefont {T.}~\bibnamefont {Prosen}},\ }\href
  {https://doi.org/10.48550/arXiv.2502.02509} {\bibinfo {title} {Fluctuations
  of stochastic charged cellular automata}} (\bibinfo {year}
  {2025}{\natexlab{b}}),\ \Eprint {https://arxiv.org/abs/2502.02509}
  {arXiv:2502.02509 [cond-mat.stat-mech]} \BibitemShut {NoStop}%
\bibitem [{\citenamefont {Moca}\ \emph
  {et~al.}(2026{\natexlab{a}})\citenamefont {Moca}, \citenamefont {P{\^a}{\c
  t}u}, \citenamefont {Zar{\'a}nd},\ and\ \citenamefont
  {D{\'o}ra}}]{MocaPatuZarandDora2026}%
  \BibitemOpen
  \bibfield  {author} {\bibinfo {author} {\bibfnamefont {C.~P.}\ \bibnamefont
  {Moca}}, \bibinfo {author} {\bibfnamefont {O.~I.}\ \bibnamefont {P{\^a}{\c
  t}u}}, \bibinfo {author} {\bibfnamefont {G.}~\bibnamefont {Zar{\'a}nd}},\
  and\ \bibinfo {author} {\bibfnamefont {B.}~\bibnamefont {D{\'o}ra}},\ }\href
  {https://doi.org/10.48550/arXiv.2609.04814} {\bibinfo {title} {Spin--charge
  subordination in the infinite-$u$ {SU(N)} hubbard chain}} (\bibinfo {year}
  {2026}{\natexlab{a}}),\ \Eprint {https://arxiv.org/abs/2609.04814}
  {arXiv:2609.04814 [cond-mat.quant-gas]} \BibitemShut {NoStop}%
\bibitem [{\citenamefont {R{\k{e}}kas}\ \emph {et~al.}(2026)\citenamefont
  {R{\k{e}}kas}, \citenamefont {Mierzejewski}, \citenamefont {Lenar{\v c}i{\v
  c}},\ and\ \citenamefont {Prelov{\v
  s}ek}}]{RekasMierzejewskiLenarcicPrelovsek2026}%
  \BibitemOpen
  \bibfield  {author} {\bibinfo {author} {\bibfnamefont {J.}~\bibnamefont
  {R{\k{e}}kas}}, \bibinfo {author} {\bibfnamefont {M.}~\bibnamefont
  {Mierzejewski}}, \bibinfo {author} {\bibfnamefont {Z.}~\bibnamefont {Lenar{\v
  c}i{\v c}}},\ and\ \bibinfo {author} {\bibfnamefont {P.}~\bibnamefont
  {Prelov{\v s}ek}},\ }\href {https://doi.org/10.48550/arXiv.2603.19743}
  {\bibinfo {title} {Spin subdiffusion in perturbed infinite-{U} {H}ubbard
  chain}} (\bibinfo {year} {2026}),\ \Eprint {https://arxiv.org/abs/2603.19743}
  {arXiv:2603.19743 [cond-mat.str-el]} \BibitemShut {NoStop}%
\bibitem [{\citenamefont {Wright}(1933)}]{Wright1933}%
  \BibitemOpen
  \bibfield  {author} {\bibinfo {author} {\bibfnamefont {E.~M.}\ \bibnamefont
  {Wright}},\ }\bibfield  {title} {\bibinfo {title} {On the coefficients of
  power series having exponential singularities},\ }\href
  {https://doi.org/10.1112/jlms/s1-8.1.71} {\bibfield  {journal} {\bibinfo
  {journal} {Journal of the London Mathematical Society}\ }\textbf {\bibinfo
  {volume} {s1-8}},\ \bibinfo {pages} {71} (\bibinfo {year}
  {1933})}\BibitemShut {NoStop}%
\bibitem [{\citenamefont {Mainardi}\ \emph {et~al.}(2010)\citenamefont
  {Mainardi}, \citenamefont {Mura},\ and\ \citenamefont
  {Pagnini}}]{Mainardi2010}%
  \BibitemOpen
  \bibfield  {author} {\bibinfo {author} {\bibfnamefont {F.}~\bibnamefont
  {Mainardi}}, \bibinfo {author} {\bibfnamefont {A.}~\bibnamefont {Mura}},\
  and\ \bibinfo {author} {\bibfnamefont {G.}~\bibnamefont {Pagnini}},\
  }\bibfield  {title} {\bibinfo {title} {The {M}-wright function in
  time-fractional diffusion processes: A tutorial survey},\ }\href
  {https://doi.org/10.1155/2010/104505} {\bibfield  {journal} {\bibinfo
  {journal} {International Journal of Differential Equations}\ }\textbf
  {\bibinfo {volume} {2010}},\ \bibinfo {pages} {104505} (\bibinfo {year}
  {2010})}\BibitemShut {NoStop}%
\bibitem [{\citenamefont {Moca}\ \emph
  {et~al.}(2026{\natexlab{b}})\citenamefont {Moca}, \citenamefont {Patu},\ and\
  \citenamefont {Dora}}]{SM}%
  \BibitemOpen
  \bibfield  {author} {\bibinfo {author} {\bibfnamefont {C.}~\bibnamefont
  {Moca}}, \bibinfo {author} {\bibfnamefont {O.}~\bibnamefont {Patu}},\ and\
  \bibinfo {author} {\bibfnamefont {B.}~\bibnamefont {Dora}},\ }\bibfield
  {title} {\bibinfo {title} {Supplemental material},\ }\href@noop {} {\
  (\bibinfo {year} {2026}{\natexlab{b}})}\BibitemShut {NoStop}%
\bibitem [{\citenamefont {Derrida}\ and\ \citenamefont
  {Gerschenfeld}(2009)}]{DG09}%
  \BibitemOpen
  \bibfield  {author} {\bibinfo {author} {\bibfnamefont {B.}~\bibnamefont
  {Derrida}}\ and\ \bibinfo {author} {\bibfnamefont {A.}~\bibnamefont
  {Gerschenfeld}},\ }\bibfield  {title} {\bibinfo {title} {Current fluctuations
  of the one dimensional symmetric simple exclusion process with step initial
  condition},\ }\href {https://doi.org/10.1007/s10955-009-9772-7} {\bibfield
  {journal} {\bibinfo  {journal} {Journal of Statistical Physics}\ }\textbf
  {\bibinfo {volume} {136}},\ \bibinfo {pages} {1} (\bibinfo {year}
  {2009})}\BibitemShut {NoStop}%
\bibitem [{\citenamefont {Bodineau}\ and\ \citenamefont
  {Derrida}(2004)}]{BodineauDerrida2004}%
  \BibitemOpen
  \bibfield  {author} {\bibinfo {author} {\bibfnamefont {T.}~\bibnamefont
  {Bodineau}}\ and\ \bibinfo {author} {\bibfnamefont {B.}~\bibnamefont
  {Derrida}},\ }\bibfield  {title} {\bibinfo {title} {Current fluctuations in
  nonequilibrium diffusive systems: An additivity principle},\ }\href
  {https://doi.org/10.1103/PhysRevLett.92.180601} {\bibfield  {journal}
  {\bibinfo  {journal} {Physical Review Letters}\ }\textbf {\bibinfo {volume}
  {92}},\ \bibinfo {pages} {180601} (\bibinfo {year} {2004})}\BibitemShut
  {NoStop}%
\bibitem [{\citenamefont {Bertini}\ \emph {et~al.}(2005)\citenamefont
  {Bertini}, \citenamefont {Sole}, \citenamefont {Gabrielli}, \citenamefont
  {Jona-Lasinio},\ and\ \citenamefont
  {Landim}}]{BertiniDeSoleGabrielliJonaLasinioLandim2005}%
  \BibitemOpen
  \bibfield  {author} {\bibinfo {author} {\bibfnamefont {L.}~\bibnamefont
  {Bertini}}, \bibinfo {author} {\bibfnamefont {A.~D.}\ \bibnamefont {Sole}},
  \bibinfo {author} {\bibfnamefont {D.}~\bibnamefont {Gabrielli}}, \bibinfo
  {author} {\bibfnamefont {G.}~\bibnamefont {Jona-Lasinio}},\ and\ \bibinfo
  {author} {\bibfnamefont {C.}~\bibnamefont {Landim}},\ }\bibfield  {title}
  {\bibinfo {title} {Current fluctuations in stochastic lattice gases},\ }\href
  {https://doi.org/10.1103/PhysRevLett.94.030601} {\bibfield  {journal}
  {\bibinfo  {journal} {Physical Review Letters}\ }\textbf {\bibinfo {volume}
  {94}},\ \bibinfo {pages} {030601} (\bibinfo {year} {2005})}\BibitemShut
  {NoStop}%
\bibitem [{\citenamefont {Appert-Rolland}\ \emph {et~al.}(2008)\citenamefont
  {Appert-Rolland}, \citenamefont {Derrida}, \citenamefont {Lecomte},\ and\
  \citenamefont {van Wijland}}]{AppertRollandDerridaLecomteVanWijland2008}%
  \BibitemOpen
  \bibfield  {author} {\bibinfo {author} {\bibfnamefont {C.}~\bibnamefont
  {Appert-Rolland}}, \bibinfo {author} {\bibfnamefont {B.}~\bibnamefont
  {Derrida}}, \bibinfo {author} {\bibfnamefont {V.}~\bibnamefont {Lecomte}},\
  and\ \bibinfo {author} {\bibfnamefont {F.}~\bibnamefont {van Wijland}},\
  }\bibfield  {title} {\bibinfo {title} {Universal cumulants of the current in
  diffusive systems on a ring},\ }\href
  {https://doi.org/10.1103/PhysRevE.78.021122} {\bibfield  {journal} {\bibinfo
  {journal} {Physical Review E}\ }\textbf {\bibinfo {volume} {78}},\ \bibinfo
  {pages} {021122} (\bibinfo {year} {2008})}\BibitemShut {NoStop}%
\bibitem [{\citenamefont {Bertini}\ \emph {et~al.}(2015)\citenamefont
  {Bertini}, \citenamefont {Sole}, \citenamefont {Gabrielli}, \citenamefont
  {Jona-Lasinio},\ and\ \citenamefont
  {Landim}}]{BertiniDeSoleGabrielliJonaLasinioLandim2015}%
  \BibitemOpen
  \bibfield  {author} {\bibinfo {author} {\bibfnamefont {L.}~\bibnamefont
  {Bertini}}, \bibinfo {author} {\bibfnamefont {A.~D.}\ \bibnamefont {Sole}},
  \bibinfo {author} {\bibfnamefont {D.}~\bibnamefont {Gabrielli}}, \bibinfo
  {author} {\bibfnamefont {G.}~\bibnamefont {Jona-Lasinio}},\ and\ \bibinfo
  {author} {\bibfnamefont {C.}~\bibnamefont {Landim}},\ }\bibfield  {title}
  {\bibinfo {title} {Macroscopic fluctuation theory},\ }\href
  {https://doi.org/10.1103/RevModPhys.87.593} {\bibfield  {journal} {\bibinfo
  {journal} {Reviews of Modern Physics}\ }\textbf {\bibinfo {volume} {87}},\
  \bibinfo {pages} {593} (\bibinfo {year} {2015})}\BibitemShut {NoStop}%
\bibitem [{\citenamefont {Haken}\ and\ \citenamefont
  {Reineker}(1972)}]{HakenReineker1972}%
  \BibitemOpen
  \bibfield  {author} {\bibinfo {author} {\bibfnamefont {H.}~\bibnamefont
  {Haken}}\ and\ \bibinfo {author} {\bibfnamefont {P.}~\bibnamefont
  {Reineker}},\ }\bibfield  {title} {\bibinfo {title} {The coupled coherent and
  incoherent motion of excitons and its influence on the line shape of optical
  absorption},\ }\href@noop {} {\bibfield  {journal} {\bibinfo  {journal}
  {Zeitschrift f{\"u}r Physik}\ }\textbf {\bibinfo {volume} {249}},\ \bibinfo
  {pages} {253} (\bibinfo {year} {1972})}\BibitemShut {NoStop}%
\bibitem [{\citenamefont {Haken}\ and\ \citenamefont
  {Strobl}(1973)}]{HakenStrobl1973}%
  \BibitemOpen
  \bibfield  {author} {\bibinfo {author} {\bibfnamefont {H.}~\bibnamefont
  {Haken}}\ and\ \bibinfo {author} {\bibfnamefont {G.}~\bibnamefont {Strobl}},\
  }\bibfield  {title} {\bibinfo {title} {An exactly solvable model for coherent
  and incoherent exciton motion},\ }\href {https://doi.org/10.1007/BF01399723}
  {\bibfield  {journal} {\bibinfo  {journal} {Zeitschrift f{\"u}r Physik}\
  }\textbf {\bibinfo {volume} {262}},\ \bibinfo {pages} {135} (\bibinfo {year}
  {1973})}\BibitemShut {NoStop}%
\bibitem [{\citenamefont {Boll}\ \emph {et~al.}(2016)\citenamefont {Boll},
  \citenamefont {Hilker}, \citenamefont {Salomon}, \citenamefont {Omran},
  \citenamefont {Nespolo}, \citenamefont {Pollet}, \citenamefont {Bloch},\ and\
  \citenamefont {Gross}}]{Boll2016}%
  \BibitemOpen
  \bibfield  {author} {\bibinfo {author} {\bibfnamefont {M.}~\bibnamefont
  {Boll}}, \bibinfo {author} {\bibfnamefont {T.~A.}\ \bibnamefont {Hilker}},
  \bibinfo {author} {\bibfnamefont {G.}~\bibnamefont {Salomon}}, \bibinfo
  {author} {\bibfnamefont {A.}~\bibnamefont {Omran}}, \bibinfo {author}
  {\bibfnamefont {J.}~\bibnamefont {Nespolo}}, \bibinfo {author} {\bibfnamefont
  {L.}~\bibnamefont {Pollet}}, \bibinfo {author} {\bibfnamefont
  {I.}~\bibnamefont {Bloch}},\ and\ \bibinfo {author} {\bibfnamefont
  {C.}~\bibnamefont {Gross}},\ }\bibfield  {title} {\bibinfo {title} {Spin- and
  density-resolved microscopy of antiferromagnetic correlations in
  {Fermi-Hubbard} chains},\ }\href {https://doi.org/10.1126/science.aag1635}
  {\bibfield  {journal} {\bibinfo  {journal} {Science}\ }\textbf {\bibinfo
  {volume} {353}},\ \bibinfo {pages} {1257} (\bibinfo {year}
  {2016})}\BibitemShut {NoStop}%
\bibitem [{\citenamefont {Hilker}\ \emph {et~al.}(2017)\citenamefont {Hilker},
  \citenamefont {Salomon}, \citenamefont {Grusdt}, \citenamefont {Omran},
  \citenamefont {Boll}, \citenamefont {Demler}, \citenamefont {Bloch},\ and\
  \citenamefont {Gross}}]{Hilker2017}%
  \BibitemOpen
  \bibfield  {author} {\bibinfo {author} {\bibfnamefont {T.~A.}\ \bibnamefont
  {Hilker}}, \bibinfo {author} {\bibfnamefont {G.}~\bibnamefont {Salomon}},
  \bibinfo {author} {\bibfnamefont {F.}~\bibnamefont {Grusdt}}, \bibinfo
  {author} {\bibfnamefont {A.}~\bibnamefont {Omran}}, \bibinfo {author}
  {\bibfnamefont {M.}~\bibnamefont {Boll}}, \bibinfo {author} {\bibfnamefont
  {E.}~\bibnamefont {Demler}}, \bibinfo {author} {\bibfnamefont
  {I.}~\bibnamefont {Bloch}},\ and\ \bibinfo {author} {\bibfnamefont
  {C.}~\bibnamefont {Gross}},\ }\bibfield  {title} {\bibinfo {title} {Revealing
  hidden antiferromagnetic correlations in doped {Hubbard} chains via string
  correlators},\ }\href {https://doi.org/10.1126/science.aam8990} {\bibfield
  {journal} {\bibinfo  {journal} {Science}\ }\textbf {\bibinfo {volume}
  {357}},\ \bibinfo {pages} {484} (\bibinfo {year} {2017})}\BibitemShut
  {NoStop}%
\bibitem [{\citenamefont {Vijayan}\ \emph {et~al.}(2020)\citenamefont
  {Vijayan}, \citenamefont {Sompet}, \citenamefont {Salomon}, \citenamefont
  {Koepsell}, \citenamefont {Hirthe}, \citenamefont {Bohrdt}, \citenamefont
  {Grusdt}, \citenamefont {Bloch},\ and\ \citenamefont {Gross}}]{Vijayan2020}%
  \BibitemOpen
  \bibfield  {author} {\bibinfo {author} {\bibfnamefont {J.}~\bibnamefont
  {Vijayan}}, \bibinfo {author} {\bibfnamefont {P.}~\bibnamefont {Sompet}},
  \bibinfo {author} {\bibfnamefont {G.}~\bibnamefont {Salomon}}, \bibinfo
  {author} {\bibfnamefont {J.}~\bibnamefont {Koepsell}}, \bibinfo {author}
  {\bibfnamefont {S.}~\bibnamefont {Hirthe}}, \bibinfo {author} {\bibfnamefont
  {A.}~\bibnamefont {Bohrdt}}, \bibinfo {author} {\bibfnamefont
  {F.}~\bibnamefont {Grusdt}}, \bibinfo {author} {\bibfnamefont
  {I.}~\bibnamefont {Bloch}},\ and\ \bibinfo {author} {\bibfnamefont
  {C.}~\bibnamefont {Gross}},\ }\bibfield  {title} {\bibinfo {title}
  {Time-resolved observation of spin-charge deconfinement in fermionic
  {Hubbard} chains},\ }\href {https://doi.org/10.1126/science.aay2354}
  {\bibfield  {journal} {\bibinfo  {journal} {Science}\ }\textbf {\bibinfo
  {volume} {367}},\ \bibinfo {pages} {186} (\bibinfo {year}
  {2020})}\BibitemShut {NoStop}%
\bibitem [{\citenamefont {Koepsell}\ \emph {et~al.}(2020)\citenamefont
  {Koepsell}, \citenamefont {Hirthe}, \citenamefont {Bourgund}, \citenamefont
  {Sompet}, \citenamefont {Vijayan}, \citenamefont {Salomon}, \citenamefont
  {Gross},\ and\ \citenamefont {Bloch}}]{Koepsell2020}%
  \BibitemOpen
  \bibfield  {author} {\bibinfo {author} {\bibfnamefont {J.}~\bibnamefont
  {Koepsell}}, \bibinfo {author} {\bibfnamefont {S.}~\bibnamefont {Hirthe}},
  \bibinfo {author} {\bibfnamefont {D.}~\bibnamefont {Bourgund}}, \bibinfo
  {author} {\bibfnamefont {P.}~\bibnamefont {Sompet}}, \bibinfo {author}
  {\bibfnamefont {J.}~\bibnamefont {Vijayan}}, \bibinfo {author} {\bibfnamefont
  {G.}~\bibnamefont {Salomon}}, \bibinfo {author} {\bibfnamefont
  {C.}~\bibnamefont {Gross}},\ and\ \bibinfo {author} {\bibfnamefont
  {I.}~\bibnamefont {Bloch}},\ }\bibfield  {title} {\bibinfo {title} {Robust
  bilayer charge pumping for spin- and density-resolved quantum gas
  microscopy},\ }\href {https://doi.org/10.1103/PhysRevLett.125.010403}
  {\bibfield  {journal} {\bibinfo  {journal} {Physical Review Letters}\
  }\textbf {\bibinfo {volume} {125}},\ \bibinfo {pages} {010403} (\bibinfo
  {year} {2020})}\BibitemShut {NoStop}%
\bibitem [{\citenamefont {Bouganne}\ \emph {et~al.}(2020)\citenamefont
  {Bouganne}, \citenamefont {{Bosch Aguilera}}, \citenamefont {Ghermaoui},
  \citenamefont {Beugnon},\ and\ \citenamefont {Gerbier}}]{Bouganne2020}%
  \BibitemOpen
  \bibfield  {author} {\bibinfo {author} {\bibfnamefont {R.}~\bibnamefont
  {Bouganne}}, \bibinfo {author} {\bibfnamefont {M.}~\bibnamefont {{Bosch
  Aguilera}}}, \bibinfo {author} {\bibfnamefont {A.}~\bibnamefont {Ghermaoui}},
  \bibinfo {author} {\bibfnamefont {J.}~\bibnamefont {Beugnon}},\ and\ \bibinfo
  {author} {\bibfnamefont {F.}~\bibnamefont {Gerbier}},\ }\bibfield  {title}
  {\bibinfo {title} {Anomalous decay of coherence in a dissipative many-body
  system},\ }\href {https://doi.org/10.1038/s41567-019-0678-2} {\bibfield
  {journal} {\bibinfo  {journal} {Nature Physics}\ }\textbf {\bibinfo {volume}
  {16}},\ \bibinfo {pages} {21} (\bibinfo {year} {2020})}\BibitemShut {NoStop}%
\end{thebibliography}%

\newpage
\setcounter{equation}{0}
\setcounter{figure}{0}
\renewcommand{\theequation}{S\arabic{equation}}
\renewcommand{\thefigure}{S\arabic{figure}}
\begin{center}
{\large\bfseries Supplemental Material}
\end{center}

\subsection{Stochastic-determinant representation}
\label{app:stochastic}
Hermitian dephasing admits a stochastic-potential representation~\cite{Daley2014,Manzano2020,Prosen2010}. We evolve with $H_\xi(t)=H_{\rm sf}+\sum_j\xi_j(t)n_j$, where $\overline{\xi_j(t)}=0$ and $\overline{\xi_j(t)\xi_k(t')}=\gamma\delta_{jk}\delta(t-t')$, interpreted in the Stratonovich sense. The evolution remains quadratic for each noise realization, allowing the finite-$\gamma$ charge FCS to be written as an average of Klich determinants,
\begin{equation}
\chi_{\subc}(\lambda,t)=
\overline{\det\!\left[(1-\bar n)\one+
\bar n e^{-i\lambda P_L}e^{i\lambda P_L(t;\xi)}\right]}.
\label{eq:stochastic_klich}
\end{equation}
The noise average is taken \emph{after} evaluating each determinant. Equation~\eqref{eq:stochastic_klich} is therefore an exact microscopic representation at arbitrary dephasing strength: free-fermion structure is retained trajectory by trajectory, while the averaged FCS describes dissipative evolution.

\subsection{Clean-chain cumulants, charge-transfer distributions, and non-Gaussian corrections}
\label{sec:clean_chain}

\begin{figure}[b]
\centering
\begin{minipage}[t]{0.48\columnwidth}
\raggedright\textbf{(a)}\\
\safeincludegraphics[width=\linewidth]{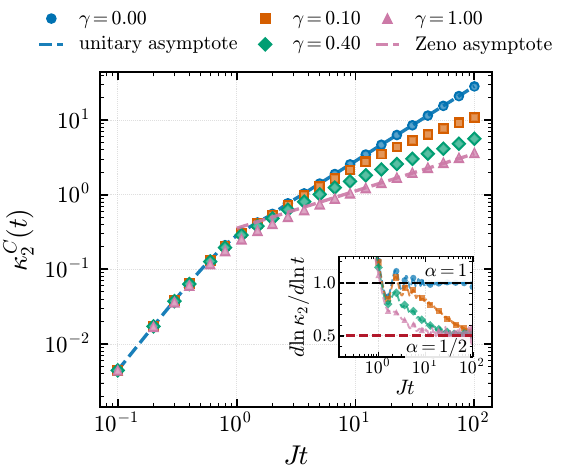}
\end{minipage}\hfill
\begin{minipage}[t]{0.48\columnwidth}
\raggedright\textbf{(b)}\\
\safeincludegraphics[width=\linewidth]{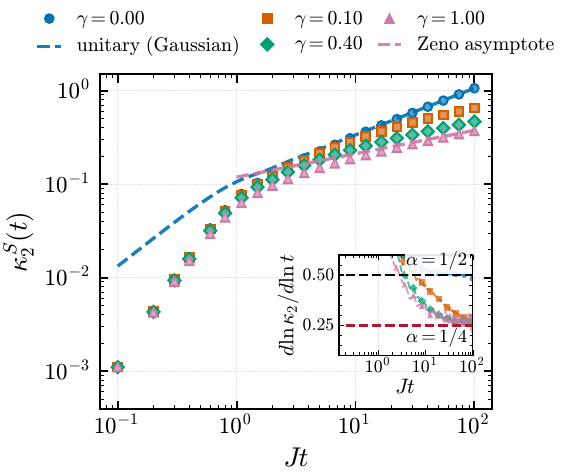}
\end{minipage}
\caption{Second cumulants across the crossover. (a) Charge: $\kappa_2^{\subc}(t,\gamma)$ grows linearly at $\gamma=0$ and crosses to $t^{1/2}$ for every $\gamma>0$; dashed: Eq.~\eqref{eq:variance_zeno}, which by Eq.~\eqref{eq:kappa2_hsr} is the exact long-time asymptote for every $\gamma>0$. (b) Spin: $\kappa_2^{\subs}(t,\gamma)$ follows Eq.~\eqref{eq:spin_variance} and crosses from $t^{1/2}$ to $t^{1/4}$, i.e., widths $t^{1/4}$ and $t^{1/8}$; dashed: unitary and Zeno asymptotes. Insets: effective exponents $d\ln\kappa_2/d\ln t$. Stochastic determinant with $L=200$ and 2000 noise realizations.}
\label{fig:cumulants}
\end{figure}

Figure~\ref{fig:cumulants} shows the second cumulants of the clean chain, $W=0$, for $\gamma=0$, $0.1J$, $0.4J$, and $J$. At $\gamma=0$ the charge variance follows Eq.~\eqref{eq:variance_unitary} over the whole time window and grows linearly for $Jt\gtrsim1$. For $\gamma>0$ the data coincide with the unitary curve at short times, depart from it near $\gamma t\approx\pi/2$, and cross over to the diffusive law $\kappa_2^{\subc}\propto t^{1/2}$: the effective exponent in the inset drops from $1$ to $1/2$. For $\gamma=J$ the data follow Eq.~\eqref{eq:variance_zeno} from $Jt\approx3$, whereas for $\gamma=0.1J$ the exponent is still slightly above $1/2$ at $Jt=100$. The spin cumulant, Fig.~\ref{fig:cumulants}(b), follows the charge through the exact identity, Eq.~\eqref{eq:second_exact}: its exponent crosses from $1/2$ to $1/4$, and at long times the data approach Eq.~\eqref{eq:spin_variance} evaluated with the unitary and Zeno forms of $\kappa_2^{\subc}$. At short times, $Jt\lesssim3$, the charge transfer is mostly $Q=0,\pm1$, so $\avg{|Q|}\simeq\kappa_2^{\subc}$ and $\kappa_2^{\subs}\simeq\kappa_2^{\subc}/4$ lies below the Gaussian estimate, Eq.~\eqref{eq:spin_variance}.

\begin{figure}[t]
\centering
\begin{minipage}[t]{0.8\columnwidth}
\raggedright\textbf{(a)}\\
\safeincludegraphics[width=\linewidth]{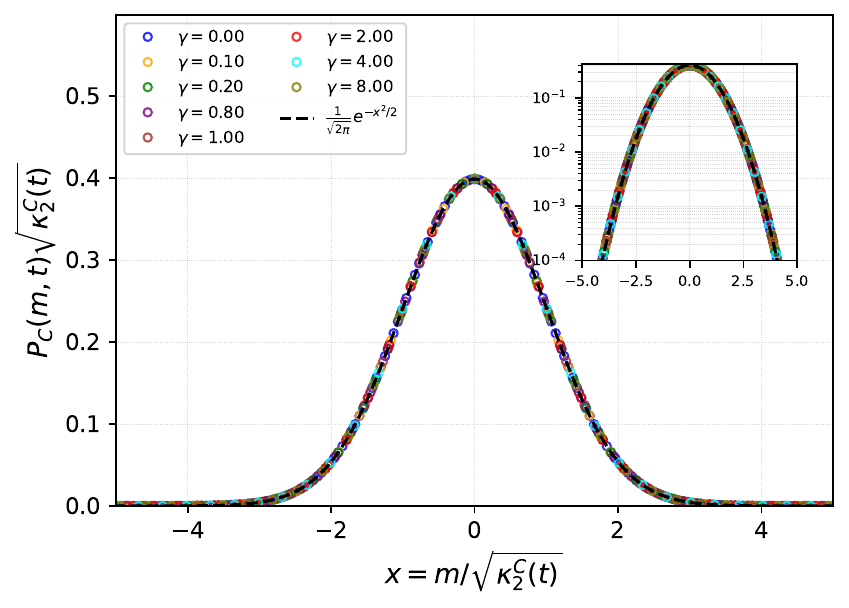}
\end{minipage}\hfill
\begin{minipage}[t]{0.8\columnwidth}
\raggedright\textbf{(b)}\\
\safeincludegraphics[width=\linewidth]{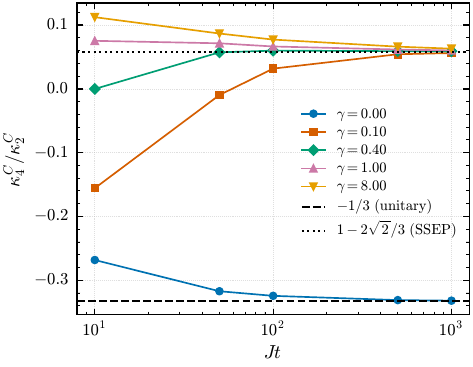}
\end{minipage}
\caption{(a) Charge-transfer distributions versus $m/\sqrt{\kappa_2^{\subc}}$. The unitary result follows from 
the exact determinant, while the finite-$\gamma$ curves approach the SSEP prediction, Eq.~\eqref{eq:ssep}.
The dashed line is the standard Gaussian, Eq.~\eqref{eq:charge_gaussian}; the non-Gaussian corrections are resolved in Fig.~\ref{fig:charge_gaussian_difference}.
The inset displays the same distributions on a logarithmic scale. (b) Cumulant ratio $\kappa_4^{\subc}/\kappa_2^{\subc}$ versus $Jt$; dashed: unitary value $-1/3$, dotted: SSEP value $1-2\sqrt2/3$, Eq.~\eqref{eq:k4k2_ssep}. Finite-$\gamma$ results use $1000$ noise realizations; system size $L=2000$.}
\label{fig:charge_distributions}
\end{figure}

Figure~\ref{fig:charge_distributions}(a) shows the charge-transfer distributions in the scaling variable $m/\sqrt{\kappa_2^{\subc}}$. For all $\gamma$ they collapse onto the Gaussian, Eq.~\eqref{eq:charge_gaussian}, down to probabilities of order $10^{-4}$ (inset), so the central charge statistics entering Eq.~\eqref{eq:mwright_scaling} are Gaussian at any dephasing strength. The non-Gaussian corrections are small on this scale; they are resolved by the fourth cumulant and by the difference from the Gaussian.

The Gaussian law, Eq.~\eqref{eq:charge_gaussian}, keeps only $\kappa_2^{\subc}$. The SSEP law, Eq.~\eqref{eq:ssep}, has all even cumulants proportional to $\sqrt{Dt}$ with fixed ratios; odd cumulants vanish at equilibrium by $m\to-m$ symmetry. Expanding Eq.~\eqref{eq:ssep} to fourth order in $\lambda$ gives, for filling $\bar n$,
\begin{equation}
\frac{\kappa_4^{\subc}}{\kappa_2^{\subc}}=1-3\sqrt2\,\bar n(1-\bar n)=1-\frac{2\sqrt2}{3}\approx0.057,
\label{eq:k4k2_ssep}
\end{equation}
whereas the unitary free-fermion chain gives $1-6\bar n(1-\bar n)=-1/3$. The two regimes therefore differ already in the sign of $\kappa_4^{\subc}$: ballistic transport gives a distribution flatter than a Gaussian, diffusive SSEP transport a slightly more peaked one.

Figure~\ref{fig:charge_distributions}(b) shows that the ratio $\kappa_4^{\subc}/\kappa_2^{\subc}$ approaches the SSEP value, Eq.~\eqref{eq:k4k2_ssep}, for every $\gamma>0$, not only for $\gamma\gg J$. Weak dephasing approaches it from below, starting from the ballistic value $-1/3$, and strong dephasing from above. The approach is slower than for $\kappa_2^{\subc}$: for $\gamma=0.1J$ it requires $Jt\sim10^2$--$10^3$, well beyond the crossover time $\pi/2\gamma$ of the variance. Together with the exact diffusion constant $D=2J^2/\gamma$ of Eq.~\eqref{eq:kappa2_hsr}, this indicates that Eq.~\eqref{eq:ssep} with $D=2J^2/\gamma$ describes the full long-time charge FCS at any dephasing strength; we have verified this for the fourth cumulant only.

\begin{figure}
  \centering
  \safeincludegraphics[width=0.95\linewidth]{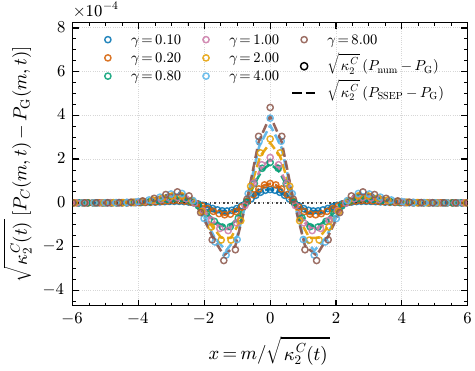}
  \caption{Deviations from Gaussianity. Difference between the numerical charge-transfer distribution and its Gaussian approximation. Dashed lines indicate the corresponding SSEP-Gaussian deviations. Stochastic determinant at $Jt=1000$ for $\gamma=0.1J$--$8J$; $L=2000$ and 1000 noise realizations.}
  \label{fig:charge_gaussian_difference}
\end{figure}

Beyond the fourth cumulant, the complete SSEP distribution retains the non-Gaussian behavior encoded in Eq.~\eqref{eq:ssep}.
Figure~\ref{fig:charge_gaussian_difference} isolates these corrections: they originate from higher charge cumulants contained in Eq.~\eqref{eq:ssep}, which are absent in the Gaussian truncation, Eq.~\eqref{eq:charge_gaussian}. The observed profile follows the SSEP benchmark (dashed curves), confirming that finite-$\gamma$ dynamics already exhibits the predicted non-Gaussian structure. 

\label{app:charge}

\subsection{Exact charge variance: derivation and crossover}
\label{app:hsr}
The second cumulant does not require the noise average. Since the initial state commutes with $N_L$ and is stationary, $\kappa_2^{\subc}=2[\avg{N_L^2}-\avg{N_L(t)N_L(0)}]$. Under the adjoint of Eq.~\eqref{eq:charge_reduced}, $f_k^\dagger f_l$ evolves with the one-particle Hamiltonian while its off-diagonal part ($k\neq l$) decays at rate $\gamma$, so $N_L(t)$ stays quadratic. Evaluating the correlator in the product initial state gives Eq.~\eqref{eq:kappa2_hsr} of the main text. For $\gamma>0$ the particle spreads coherently from a site between position-resetting events that occur at rate $\gamma$; this renewal structure gives the Fourier--Laplace transform ($r\to\theta$, $t\to z$)
\begin{equation}
\tilde p(\theta,z)=\left[\sqrt{(z+\gamma)^2+16J^2\sin^2(\theta/2)}-\gamma\right]^{-1},
\label{eq:hsr_laplace}
\end{equation}
and hence $\avg{X^2(t)}=(4J^2/\gamma^2)(\gamma t-1+e^{-\gamma t})$. For $\gamma t\gg1$, $X(t)$ is Gaussian with variance $4J^2t/\gamma$, so $\avg{|X(t)|}\simeq(8J^2t/\pi\gamma)^{1/2}$ and Eq.~\eqref{eq:kappa2_hsr} reproduces Eq.~\eqref{eq:variance_zeno} for every $\gamma>0$, not only for $\gamma\gg J$. The diffusion constant $D=2J^2/\gamma$ agrees with the exact steady state of the dephased XX chain~\cite{Znidaric2010}. For $\gamma\ll J$, $\avg{|X(t)|}=(J/\gamma)\,\Phi(\gamma t)$ with a parameter-free function $\Phi(\tau)$ that interpolates between $4\tau/\pi$ and $(8\tau/\pi)^{1/2}$; the two limits cross at $\gamma t=\pi/2$. We checked Eq.~\eqref{eq:kappa2_hsr} against exact Lindblad evolution of a six-site chain, with relative agreement better than $10^{-10}$.

\subsection{Quantum-Zeno limit in the spin sector}
\label{app:zeno}

In the Zeno regime, the full charge FCS can be retained rather than first reduced to its Gaussian central sector. Substituting Eq.~\eqref{eq:ssep} 
into the exact kernel, Eq.~\eqref{eq:poisson_kernel}, gives
\begin{equation}
\begin{aligned}
\chi_{\subs}^{\rm Z}(\lambda,t)\simeq \int_{-\pi}^{\pi}\frac{d\theta}{2\pi}\,
&\frac{\sin^2(\lambda/2)}{1-2\cos(\lambda/2)\cos\theta+\cos^2(\lambda/2)}
\\
&\times\exp\!\left\{\sqrt{Dt}\,F\!\left[-\frac89\sin^2\left(\frac{\theta}{2}\right)\right]\right\}.
\end{aligned}
\label{eq:zeno_spin_exact}
\end{equation}
Equation~\eqref{eq:zeno_spin_exact} retains the full non-Gaussian SSEP charge statistics. It also makes the exact structure explicit: frozen spin order 
fixes the kernel, while the environment enters only through the charge factor. Its small-$\lambda$ central sector reduces to Eq.~\eqref{eq:spin_variance} 
and recovers the M-Wright scaling form, Eq.~\eqref{eq:mwright_scaling}; away from that sector, it also contains the SSEP corrections.
At long times, $J^2t\gg\gamma$, the charge factor in Eq.~\eqref{eq:zeno_spin_exact} is sharply peaked at $\theta=0$ with a Gaussian profile, which again yields the M-Wright distribution. The higher-order terms of $F$ in $\theta$ remain in the exponent and modify the M-Wright tails.

\end{document}